\documentclass[twocolumn]{pasj01}

\makeatletter
\@namedef{ver@amsmath.sty}{}
\makeatother
\usepackage{amstext}

\Received{$\langle$reception date$\rangle$}
\Accepted{$\langle$acception date$\rangle$}
\Published{$\langle$publication date$\rangle$}

\newcommand{\Mstar}{M_\star}

\usepackage[switch,mathlines]{lineno}

\begin{document}

\title{The Dependence of Galaxy Properties on the Underlying 3D Matter Density Field at $2.0<z<2.5$}

\author{Rieko \textsc{Momose}\altaffilmark{1,2,3,4}}
\altaffiltext{1}{Observatories of the Carnegie Institution for Science, 813 Santa Barbara Street, Pasadena, CA 91101, USA}
\altaffiltext{2}{Department of Astronomy, School of Science, The University of Tokyo, 7-3-1 Hongo, Bunkyo-ku, Tokyo, 113-0033, Japan}
\altaffiltext{3}{Kavli Institute for the Physics and Mathematics of the Universe (WPI), UTIAS, The University of Tokyo, Kashiwa, Chiba 277-8583, Japan}
\altaffiltext{4}{The Institute for AI and Beyond, The University of Tokyo, Tokyo 113-8655, Japan}
\email{rmomose@carnegiescience.edu}

\author{Khee-Gan \textsc{Lee}\altaffilmark{3}}

\author{Metin \textsc{Ata}\altaffilmark{7,3}}
\altaffiltext{7}{The Oskar Klein Centre, Department of Physics, Stockholm University, AlbaNova University Centre, SE 106 91 Stockholm, Sweden}

\author{Benjamin \textsc{Horowitz}\altaffilmark{5,6}}
\altaffiltext{5}{Department of Astrophysical Sciences, Princeton University, Princeton, NJ 08544, USA}
\altaffiltext{6}{Lawrence Berkeley National Laboratory, 1 Cyclotron Road, Berkeley, CA 94720, USA}

\author{Jeyhan S. \textsc{Kartaltepe}\altaffilmark{8}}
\altaffiltext{8}{Laboratory for Multiwavelength Astrophysics, School of Physics and Astronomy, Rochester Institute of Technology, 84 Lomb Memorial Drive, Rochester, NY 14623, USA}


\maketitle

\begin{abstract}
We study the environmental effect of galaxy evolution as a function of the underlying 3D dark matter density for the first time at $z=2-2.5$, in which the underlying matter density is reconstructed from observed galaxies through dynamical forward modeling techniques. 
Utilizing this map, we investigate the dependence of the star formation activities and galaxy types (mergers, submillimeter galaxies, active galactic nuclei, and quiescent galaxies) on the matter overdensity $\Delta_\text{local}$ and stellar mass. 
For the first time, we are able to probe underdense regions ($\Delta_\text{local}<1$) in addition to overdensities.
We find that star formation activity generally depends on the stellar mass, not the matter density. 
We also find evidence that:
(1) an absence of mergers and submillimeter galaxies in higher-density regions but otherwise no trend across lower-density bins, (2) the increase of active galactic nuclei and quiescent galaxy prevalence as a function of matter density, and (3) the increase of all aforementioned categories with the stellar mass. 
These results indicate that stellar mass is the main driver of galaxy evolution at the cosmic noon. 
Our novel approach directly using reconstructed dark matter density maps demonstrates the new capability of the environmental effect studies in galaxy evolution at higher redshift.  
\end{abstract}


\section{Introduction} \label{sec:intro}
The understanding of galaxy formation and evolution has accelerated for over the last decades thanks to numerous galaxy observations from local to the high-$z$ universe. 
One of the notable findings is the bimodality of galaxy populations in color--magnitude or color--star formation rate (SFR) / specific SFR (sSFR) diagrams in the local to as far as redshift $z\sim3$ universe (e.g., \cite{Strateva01,Kauffmann03,Blanton03,Blanton05,Baldry04,Willmer06,Wyder07,Williams09,Blanton09,Brammer09,Wetzel12}). Previous studies have found two distinct galaxy populations within the bimodality. One is referred to as the blue cloud consisting of star-forming galaxies, and the other is the red sequence of early-type galaxies with low or absence of new star formation. This galaxy bimodality is interpreted as the consequence of a change in two galaxy populations transitioning from the blue star-forming to the red quiescent over time. In other words, star-forming galaxies transform their physical properties, such as shape, color, and stellar ages, while quenching their star formation. Understanding the physical origin or processes that halt star formation and transit from the star-forming to quiescent phase in galaxies is therefore fundamental to unveiling galaxy formation and evolution. 

Although several physical mechanisms for quenching star formation in galaxies have been proposed in the literature, they can generally be categorized as ``internal'' and ``external processes'' (e.g., \cite{Peng10}). The ``internal process'' or ``mass quenching'' includes any processes operating in individual galaxies themselves that link to or depend on the galaxy mass, such as feedback from active galactic nuclei (AGN) from the central supermassive black hole (SMBH) (e.g., \cite{Croton06,Fabian12,Fang13,Ito22}), supernova explosions, and stellar wind (e.g., \cite{Dekel86,DallaVecchia08}). Those processes can scatter out or heat the cold, dense gas available for forming stars, eventually obstructing the cycle of next star formation.  

The second is an ``external process,'' also dubbed ``environmental quenching.''
This includes all physical processes originating from environments around galaxies leading to quench the star formation and work more effectively in higher density environments, such as galaxy groups, subclusters, and clusters. A variety of physical processes has been proposed in the literature.
Galaxy interactions and mergers are considered to enhance star formation while simultaneously increasing the mass of the central SMBHs that, eventually, bring on the quenching of star formation (e.g., \cite{Sanders96,Springel05a,Springel05b,Hopkins06,Hopkins08,Yuan10,Alexander12}). 
Likewise, high-speed fly-by galaxy encounters within clusters and subclusters, known as galaxy harassment, can eventually lead to the quenching while causing morphological transformations and starbursts (e.g., \cite{Moore96,Moore98,Smith10,Smith13,Smith15}). 
The removal of gas in the interstellar regions by ram pressure stripping and tidal stripping in clusters or subclusters results in strangulation, which starves the galaxy and subsequently quenches the future star formation (e.g., \cite{Gunn72,Toomre72,Larson80,Balogh00,Kawata08,McCarthy08,Bekki14,Smith16,Boselli22}). 

The dominant process for inducing galaxy quenching, either mass or environmental, is still hotly debated, though evidence indicates both processes contribute. 
It is commonly believed that at lower redshifts ($z<1$), mass quenching seems to be dominant in high mass or central galaxies, whereas the environmental quenching becomes more important in low mass or satellite galaxies (e.g., \cite{Peng10,Peng12,Sobral11}). 
This trend seems to change in the earlier universe, but the evidence is still inconclusive.
\cite{Darvish16} have studied the effect of mass and environment on galaxy properties up to $z=3$ using photometric redshift samples, and suggested that mass quenching plays a dominant role at $z>1$ and the environmental quenching is relevant only at $z<1$. 
Several other studies have also reported similar results (e.g., \cite{Muzzin12,Nantais16}) at $z\sim 1-1.5$. 
However, environmental quenching has been observed up to $z=2-3$ in other works.
(e.g., \cite{Fossati17,Guo17,Kawinwanichakij17,Ji18,Zavala19,Chartab20,Ando20}). 
Further studies on the environmental effects of galaxy formation and evolution are necessary for deeper insight into the quenching mechanisms at higher redshift beyond $z\geq1$. 

One of the common challenges in studies of the environmental effects is correctly estimating the environment or overdensity around galaxies. They are generally assessed by galaxy distributions, often based on photometric redshifts, using different estimators, such as \textit{N}-th nearest neighbors, Friends-of-Friends grouping algorithm, counts-in-cylinders, and adaptive weighted kernel smoothing. 
However, evaluating galaxy overdensities at higher redshift ($z>2$) posses significant challenges.
Firstly, the larger uncertainties of photo-$z$ estimates at higher redshift reduce the contrast of dark matter density over tens of Megaparsecs along the line of sight. This effect gives a smaller value to galaxy overdensities than the real ones. 
Secondly, an estimated galaxy environment could be biased toward regions where higher mass galaxies mainly reside because the stellar-mass completeness limit of a survey increases with redshift. 
Thirdly, recent observations have shown that galaxy distributions cannot always trace the underlying matter density when a particular type of galaxy is used such as Ly$\alpha$ emitters (e.g., \cite{Overzier08,Toshikawa16,Shimakawa17,Shi19,Liang21,Momose21c,Momose21a,Ito21,Huang22}). One might derive incorrect overdensities if we use that type of galaxies to evaluate the environment.
Fourthly, with many traditional environmental measures such as \textit{N}-th nearest neighbors and counts-in-cylinders, it is difficult to correct for the observational selection functions even with spectroscopic data. 
Those reasons lead us to consider another methodology for evaluating the environments around galaxies at higher redshift. 

In this study, we estimate the galaxy environments based on the underlying 3D dark matter density. Our novel approach corrects for the issues mentioned above, that could arise from evaluating the dark matter distribution based on a density field derived from the spatial distribution of galaxies.
Theoretical work over the past decade has led to some novel computational techniques to reconstruct the underlying 3D matter density within a survey volume using spectroscopic galaxy positions as tracers (e.g., \cite{Jasche13,Kitaura21,Horowitz19,Horowitz21,Ata21,Ata22}). 
Those reconstructed density fields have advantages in identifying the matter density accurately and are superior to studies referring to galaxy distributions alone because they have corrected for galaxy bias, survey selection functions, and redshift space distortions.
We utilize such reconstructed matter density fields as a proxy of the environments in order to investigate the dependence of star formation activities and galaxy types on the mass and environments of galaxies at $z\sim2$. This allows us to address the dominant physical mechanisms for quenching star formation in galaxies. 

This study investigates correlations between the matter density and galaxy properties, specifically, star formation activities and the fraction of submillimeter galaxies, mergers, queiscent galaxies, and AGN. 
The layout of this paper is as follows. We introduce the data and method used in this study in Section~\ref{sec:data} and show our results in Section~\ref{sec:results}. Discussions about the galaxy evolution based on our results are presented in Section~\ref{sec:dis}.

\section{Data}
\label{sec:data}

\subsection{Matter Density Maps} 
\label{sec:tardis}

This study uses a reconstructed 3D matter density field, named the COnstrained Simulations of The COsmos field (COSTCO, \cite{Ata22}), to evaluate the local matter density around galaxies in the Cosmic Evolution Survey field (COSMOS; \cite{Capak07,Scoville07}). Although the detail of the methodology is fully described in \citet{Ata22}, we provide its summary below.  

The COSTCO is a suite of constrained simulations designed to resemble three-dimensional galaxy distributions in the COSMOS field at $2.0\leq z \leq 2.52$. Firstly, a binned galaxy density field was calculated from  a compilation of galaxy spectroscopic surveys: zCOSMOS-Deep \citep{Lilly07}, VUDS \citep{LeFevre15}, MOSDEF \citep{Kriek15}, 
and ZFIRE \citep{Nanayakkara16}, with the respective angular and radial selections carefully accounted for \citep{Ata21}. 
Using the COSMIC BIRTH algorithm, a nested Bayesian inference algorithm \citep{Kitaura21,Ata21}, initial fluctuations (at $z=100$) were estimated such that will eventually evolve into a matter density field consistent with the observational galaxy field at $2.0<z<2.55$.  A suite of cosmological \textit{N}-body simulations were then performed from the posterior sample of these initial conditions in order to describe the evolution of observed structures over cosmic time, which is dubbed COSTCO. The COSTCO matter density field used in this study covers the central 0.9 deg$^2$ region of the COSMOS footprint where most of the spectroscopic data reside, spanning a redshift range of $z=2-2.52$ and has a binning of $0.5~h^{-1}$~cMpc. 
\citet{Ata22} have estimated the effective reconstruction scale of COSTCO to be approximately
$\approx4.5~h^{-1}$~cMpc.

\begin{figure}[t]
\includegraphics[width=0.95\linewidth]{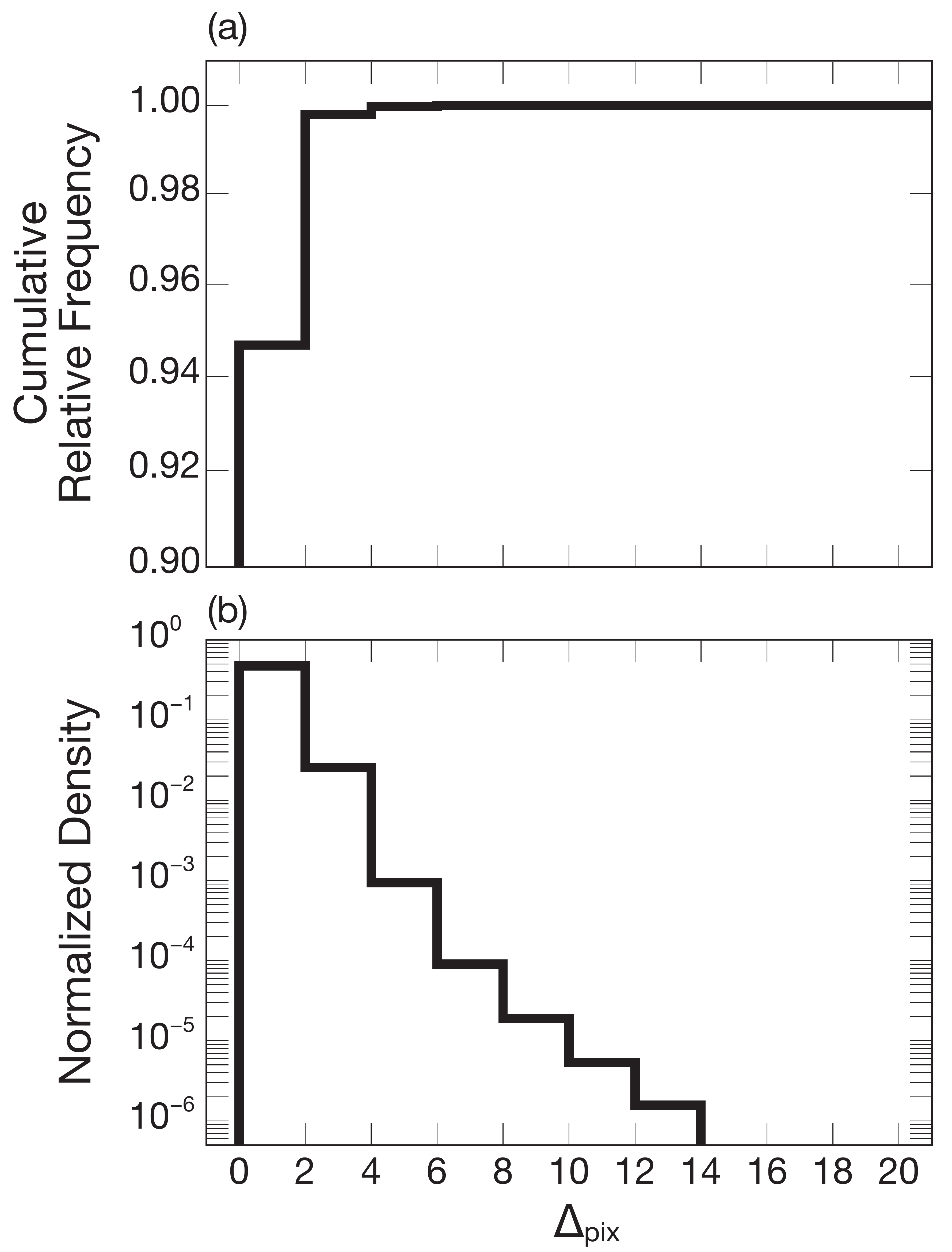}
\caption{
(a) The normalized cumulative profile and (b) number density of the density contrast, $\Delta_\text{pix}$, in the COSTCO field. 
}
\label{fig:hist_dm}
\end{figure}

Figure~\ref{fig:hist_dm} demonstrates the normalized cumulative profile and number density of the matter density contrast or ``overdensity'', $\Delta_\text{pix} \equiv \rho / \langle \rho \rangle$, in the entire COSTCO field. This $\Delta_\text{pix}$ is the output of reconstructions and one-pixel values.
The COSTCO density field is occupied by the lowest densities ($0\leq\Delta_\text{pix}<2$) with more than $90\%$ volume filling factor. 

Instead of utilizing $\Delta_\text{pix}$, we used the density contrast within $2~h^{-1}$~cMpc in a radius (i.e., $4~h^{-1}$~cMpc in a diameter) around individual galaxies as their local environment, which refers to $\Delta_\text{local}$, in the following analysis. 
We choose the scale since
[a] the smoothed simulation of $4~h^{-1}$~cMpc still showed good agreements on the mean power spectrum of the COSTCO field (see Supplementary Fig. 5 in \cite{Ata22}) and
[b] the sample size on the matter density subsamples introduced in Section \ref{sec:galaxy_all} does not vary by $4~h^{-1}$~cMpc or $4.5~h^{-1}$~cMpc.

%

\subsection{Galaxy Sample} 
\label{sec:galaxy_all}

%
\begin{figure}[t!]
\includegraphics[width=\linewidth]{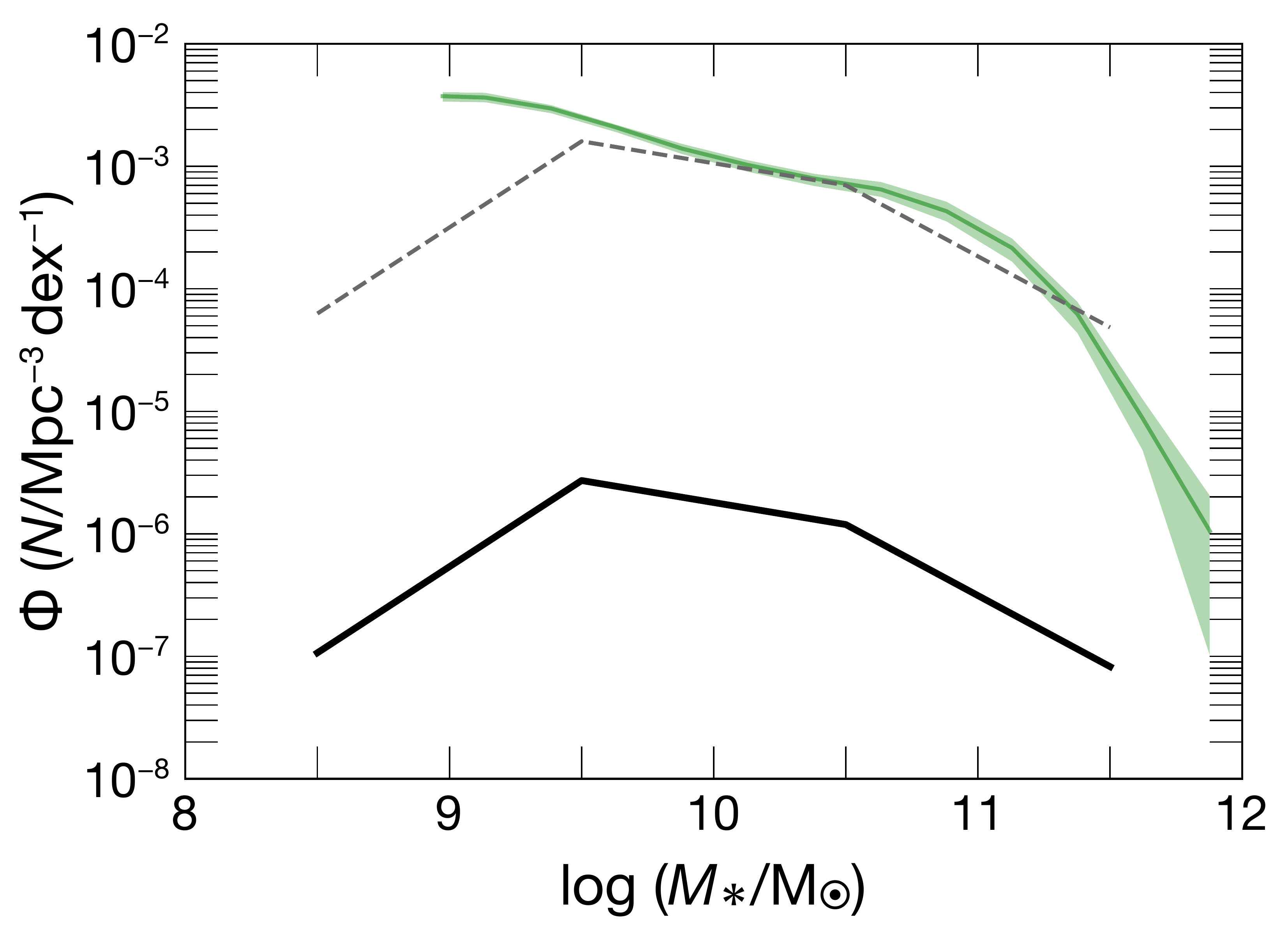}
\caption{
The stellar mass function of our mass-complete sample (black straight line) and photo-$z$ galaxies at $2.0\leq z \leq2.5$ in \citet{Weaver23} (green line). The dashed line is the mass-complete sample with offsets for shape comparison with the one by \citet{Weaver23}. 
}
\label{fig:mass_func}
\end{figure}

\begin{figure*}[t!]
\includegraphics[width=0.95\linewidth]{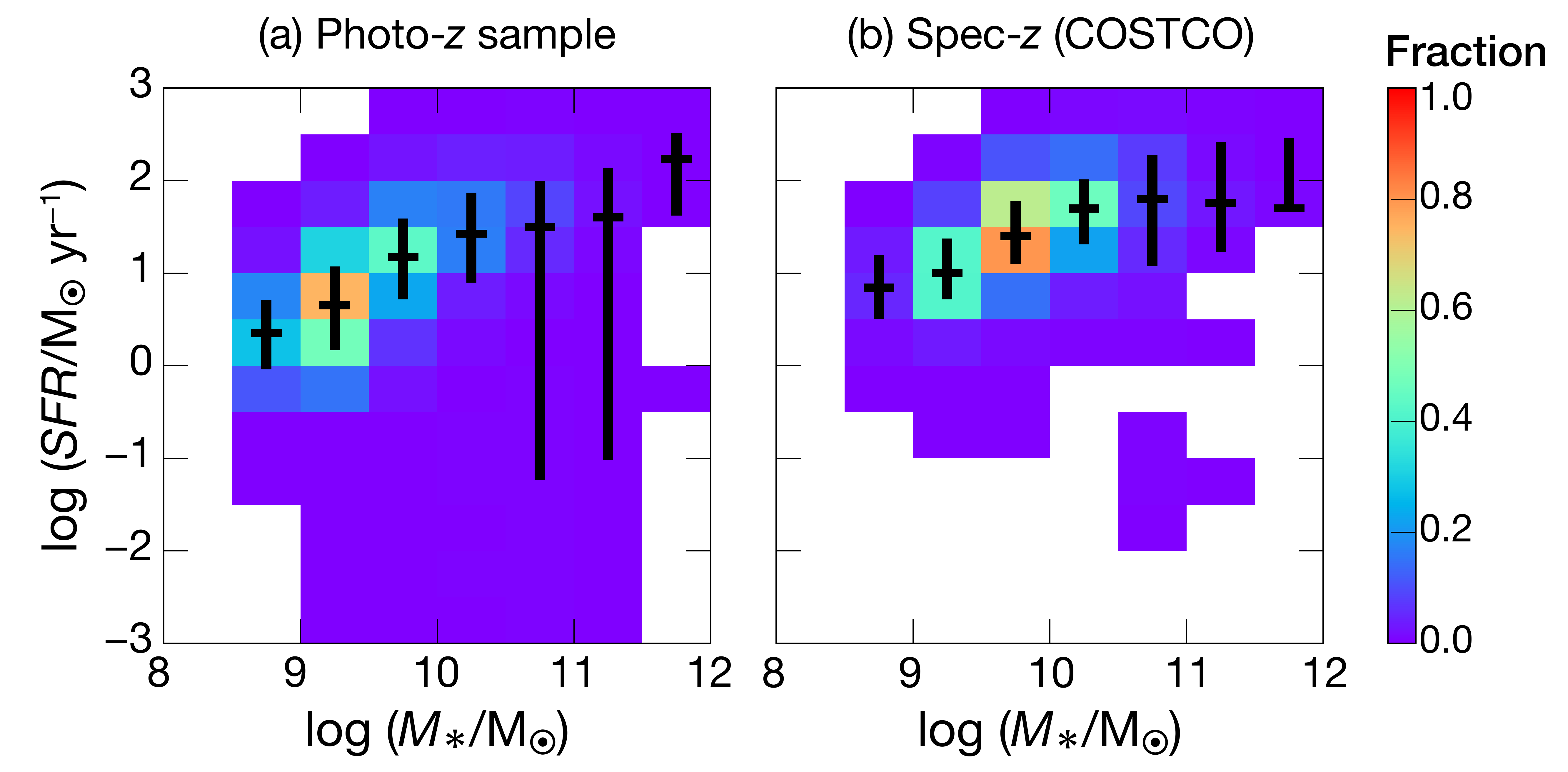}
\caption{
The heatmap of galaxies in the main sequence diagram of (a) the COSMOS2020 at $1.9\leq z \leq2.6$ and (b) mass-completeness sample in the COSTCO field. The number of galaxies in each bin is normalized. A pixel with 0 is denoted by white. The median of each $\Mstar$ pixel and the percentail 16th and 84th of \textit{SFR} is plotted by plus marks. 
}
\label{fig:heat_sample}
\end{figure*}

\begin{figure*}[t!]
\includegraphics[width=\linewidth]{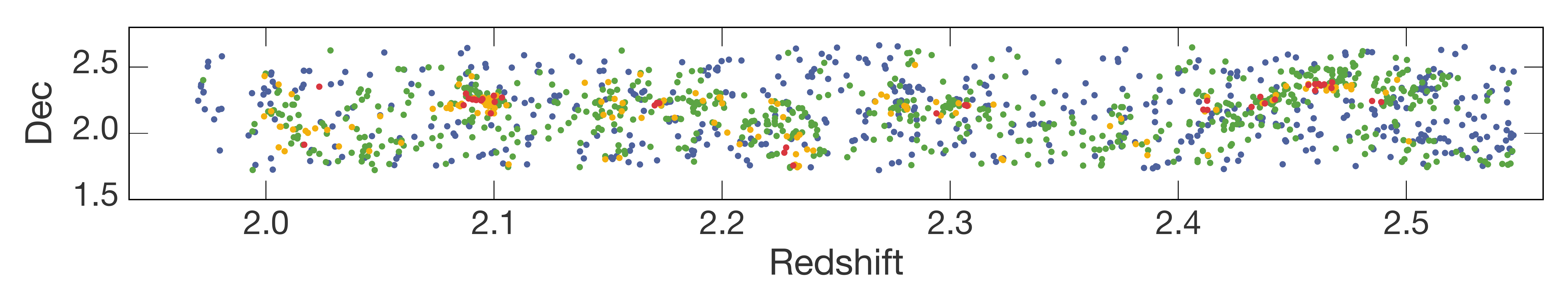}
\caption{
Distributions of our galaxy samples in the COSTCO field in the Dec-\textit{z}$_\text{spec}$ plane. Color is based on the matter density subsamples -- blue, green, yellow, and red indicate subsamples in $\Delta_\text{local}<1$, $1\leq\Delta_\text{local}<2$, $2\leq\Delta_\text{local}<4$, and $\Delta_\text{local}\geq4$. 
}
\label{fig:map_Dec_z_plane}
\end{figure*}
\begin{table}
  \tbl{The Number of Galaxies Used in This Study}{%
    \begin{tabular}{ccc}
    \hline
    (1) & (2) & (3)  \\ 
    \hline\hline
    All & 1344 & (1223, 121)  \\
    \hline
    Lowest-$\Delta$ ($\Delta_\text{local}<1$)            & 519 & (484, 35) \\
    Low-$\Delta$ ($1\leq\Delta_\text{local}<2$)          & 653 & (588, 65) \\
    Intermediate-$\Delta$ ($2\leq\Delta_\text{local}<4$) & 154 & (135, 19)  \\
    High-$\Delta$ ($\Delta_\text{local}\geq4$)           & 18 & (16, 2)   \\
    \hline
    Mergers$^\dagger$ & 19 & (9, 10) \\
    SMGs        & 11 & (1, 10) \\
    AGNs        & 23 & (4, 19) \\ 
    QGs         & 12 & (2, 10) \\
    \hline
    \end{tabular}}
  \label{tab:galaxy_type}
  \begin{tabnote}
    The number of the mass-completeness sample in the COSTCO field.
    (1) Category of subsamples. (2) Total number of galaxies in each subsample. (3) Number of low- and high-mass galaxies in a subsample. This study regards galaxies with $\Mstar<10^{10.5}$~M$_\odot$ as low-mass and with $\Mstar\geq10^{10.5}$~M$_\odot$ as high-mass, respectively. Their numbers are denoted in the table as (low-mass galaxies, high-mass galaxies). \\
    $^\dagger$ We list the number of individual merging galaxies instead of the one of merging pairs
  \end{tabnote}
\end{table}

In this study, we used several spectroscopic catalogs targeting $2.0\leq z \leq 2.55$ galaxies and constructed a complied catalog (\cite{Smolcic12,Hashimoto13,Nakajima13,Shibuya14,LeFevre15,Kriek15,Nanayakkara16,Lee16,Lee18,Brisbin17,Michalowski17,Champagne21}; Lilly et al. in prep).
First, we extracted galaxies at $2.0\leq z \leq 2.55$ laying within the COSMOS field. Then, we applied quality cuts defined in each catalog and cross-matched each other with a maximum allowable separation of $0\farcs5$. The final compiled catalog contains $1784$ galaxies, which are referred to as the ``full sample''.

The stellar mass ($\Mstar$), \textit{SFR}, and \textit{sSFR} of galaxies in the compiled catalog were taken from the latest photo-$z$ catalog, COSMOS2020 \citep{Weaver21}. 
In this study, we use the above quantities using estimates derived from spectral energy distribution (SED) fitting code, \texttt{LePhare} \citep{Ilbert06}, in \texttt{FARMER} catalog of the COSMOS2020. 
We first cross-matched the compiled spectroscopic catalog with the COSMOS2020 within a $1.0\arcsec$ radius and obtained $1428$ galaxies. 
We further applied the mass completeness criteria defined in the COSMOS2020\footnote{This equation is the same as one in Sec.6.2 of \cite{Weaver21}.},
\begin{equation}
    \frac{M_\text{lim}(z)}{\text{M}_\odot} = -1.51\times10^6(1+z) + 6.81\times10^7(1+z)^2
\end{equation}
where $M_\text{lim}$ is a redshift-dependent threshold in stellar mass (M$_\odot$). That results in $1344$ complete mass samples for the COSTCO field. Galaxies above the mass completeness criteria are denoted as ``mass-complete sample", which we used in this study.

Galaxies targeted for spectroscopic observations are generally selected from photometric surveys, implying that they are just a subset of the galaxies within the photometry catalogs. Here, we compare $\Mstar$ and \textit{SFR} of our mass-complete sample and parent photometric catalog of COSMOS2020 in order to briefly examine if we can regard our galaxy samples as a representative of the COSMOS2020 in terms of their physical properties.
In Figure \ref{fig:mass_func}, we compare the stellar mass function (SMF) of the mass-completeness sample in this study to the parent photometric catalog of COSMOS2020 \citep{Weaver23}. The black line shows our SMF, whereas the dashed line represents the normalized ones at $\Mstar=10^{10}$~M$_\odot$ for its shape comparison purposes. Note that we do not evaluate uncertainty for our SMF because we focus on its shape comparison. Besides, the bin size is larger than the one in \citet{Weaver23} due to our limited sample size. We apply the bin size of $\Mstar=10^{1}$~M$_\odot$ from $\Mstar=10^{8}$ to $10^{12}$~M$_\odot$ which is also used in our investigations shown in Section~\ref{sec:result_mass}.
Overall, our SMF has a smaller galaxy number density $\Phi(\Mstar)$ by $3-4$ orders of magnitudes than the parent photo-\textit{z} sample. The normalized SMF shown by the dashed line appears to have a consistent shape to the parent SMF of \citet{Weaver23} at $\Mstar>10^{9}$~M$_\odot$ within the uncertainties, despite the coarseness of our stellar mass bin.
We also compare the 2D distribution (heatmap) of galaxies in the star formation main sequence diagram of COSMOS2020 and this study Figure~\ref{fig:heat_sample}.
The heatmap for COSTCO shows the lack of galaxies with a lower \textit{SFR} overall $\Mstar$ range compared to samples in COSMOS2020. Besides, the peaks of both $\Mstar$ and \textit{SFR} in our samples are shifted to $0.5$ dex higher values than the COSMOS2020 samples. Nonetheless, the median values of parent samples and ours are within the uncertainties.
To sum up the comparison between parent COSMOS2020 samples and ours, our galaxies can be regarded as a proxy of their stellar mass, at least $\Mstar\geq10^9$~M$_\odot$ under the coarse $\Mstar$ bin. We also anticipate that the \textit{SFR} values should be consistent within the error, at least in each $\Mstar$ bin.

\subsubsection{Subsamples}

For this study, we construct three different subsamples to establish the link between galaxy evolution and environment or mass. The first is mass subsamples, which are used to examine the dependence on the density contrast. We divide our samples into two, low-mass and high-mass, according to their stellar masses. We define the threshold for the high-mass at $\Mstar=10^{10.5}$~M$_\odot$. 
We adopt this threshold because $\Mstar=10^{10.5}$~M$_\odot$ corresponds to the stellar mass reaching the peak of the stellar-to-halo mass relation and could be boundary mass resulting in a significant difference in the star formation history and subsequent quenching (e.g., \cite{Moster13,Behroozi13,Behroozi19}).
Note that you can find the impact of the threshold mass on the results in Appendix~\ref{sec:app_dDM_Ms}. 
The number of galaxies classified as either low-mass or high-mass is listed in Table~\ref{tab:galaxy_type}. The fraction of high-mass galaxies is about $10\%$ of the total in the COSTCO volume.

The second is the matter density subsamples to investigate mass dependence.
Our mass-complete samples are divided into four density bins based on their $\Delta_\text{local}$, where the underdense ($\Delta_\text{local}<1$), low- ($1\leq\Delta_\text{local}<2$), intermediate- ($2\leq\Delta_\text{local}<4$), and high-density contrasts ($\Delta_\text{local}\geq4$). 
The highest-densities of $\Delta_\text{local}\geq4$ typically corresponds to the central regions of protoclusters (when compared with, e.g., \cite{Ata22}), while $2\leq\Delta_\text{local}<4$ are associated with protocluster outskirts. 
Galaxies residing in $1\leq\Delta_\text{local}<2$ regions, meanwhile, can be roughly regarded as `field' galaxies. Finally, $\Delta_\text{local}<1$ are in underdensities, although we have not made a careful comparison with the cosmic void catalog of \citet{Krolewski18}.
Figure~\ref{fig:map_Dec_z_plane} shows galaxy distributions of mass-complete samples in Dec-\textit{z}$_\text{spec}$ plane with color-coded in each matter density subsample. Some of the high-density subsample galaxies are found at about the same position as well-known galaxy protoclusters -- one is found at $z\sim2.1$, and the other is at $z\sim2.44$. 
It is also worth discussing the volume fraction of each matter density subsample relative to the entire volume and the density of galaxies in that density bin. We refer to a pixel value, $\Delta_\text{pix}$, for this analysis.
Each density subsample of $\Delta_\text{pix}<1$, $1\leq\Delta_\text{pix}<2$, $2\leq\Delta_\text{pix}<4$, and $\Delta_\text{pix}\geq4$ occupies $58.2$, $36.5$, $5.1$, and $0.2\%$ in the COSTCO volume. More than half of COSTCO volume is filled with underdense region on $\Delta_\text{pix}<1$. 
Meanwhile, a number density of galaxies is ($2.8$, $5.4$, $9.6$, $27.6$)$\times10^{-6}$~cMpc$^{-3}$ in (lowest, low, intermediate, high)-density subsample. Given the mean number density for the entire COSTCO volume, $4.1\times10^{-6}$~cMpc$^{-3}$, a growth rate on a number density of galaxies with matter density is larger than the one expected from the change of matter density contrast. Note that this number density of galaxies does not vastly change when we even refer to $\Delta_\text{local}$.

Finally, we also assess the dependence of
quiescent galaxies (QGs), merging galaxies, submillimeter galaxies (SMGs) and AGN fractions on the density contrast.
The QGs are identified with the same method proposed by \citet{Ilbert13}. They have classified QGs and star-forming galaxies based on the rest frame NUV, \textit{r}, and \textit{J} colors, such that QGs are defined with
\begin{equation}
        m_\text{NUV} - m_r > 3(m_r - m_J) + 1 \ \ \rm{and} \\
        m_\text{NUV} - m_r > 3.1 
\end{equation}
Here, $m_\text{NUV}$, $m_r$, and $m_J$ are the rest-frame NUV, \textit{r}, and \textit{J} magnitudes given in the COSMOS2020 catalog. We have $12$ QGs in the COSTCO volume.   

For the merging galaxies, SMGs, and AGNs,
we performed a cross-match between the compiled catalog and several galaxy catalogs in the literature using a matching radius of $1.0\arcsec$.
To find merging galaxies, we used two different merger catalogs from \citet{Silva18} and \citet{Shah20}; details on the methodology of galaxy pairs identification can be found in the respective references. Although they have used different criteria for identifying galaxy pairs, both studies selected galaxies greater than $\Mstar>10^{10}$ M$_\odot$. We adopted the same threshold for our merger samples. 
As for SMGs, we refer to four catalogs \citep{Smolcic12,Brisbin17,Michalowski17,Champagne21} and cross-matched them with our compiled spectroscopic catalog.
For AGNs, we cross-matched with \citet{Straatman16}, \citet{Cowley16} and \citet{Delvecchio17}. \citet{Cowley16} have selected AGNs based on the rest frame radio, X-ray, or infrared luminosities. \citet{Delvecchio17}, meanwhile, have classified AGNs from 3 GHz radio sources using SED fitting, X-ray and mid-infrared diagnostics, and radio emission excess. The detailed criteria for AGN candidates is found in those studies.
The numbers of mergers, SMGs, and AGNs used in this study are given in Table~\ref{tab:galaxy_type}. Note for mergers that we count the number of individual merging galaxies, not each merging pair as one object. Because not all galaxies in merging pairs were cross-matched with the full sample, we got odd numbers.

\section{Results}
\label{sec:results}

This section shows results to investigate two dependences of galaxy properties on local density environment, $\Delta_\text{local}$, and stellar mass $\Mstar$.

\subsection{Dependence On Matter Density Contrast}

\subsubsection{Global Trend}
\label{sec:results_delta_global}

%
\begin{figure}[t]
\includegraphics[width=\linewidth]{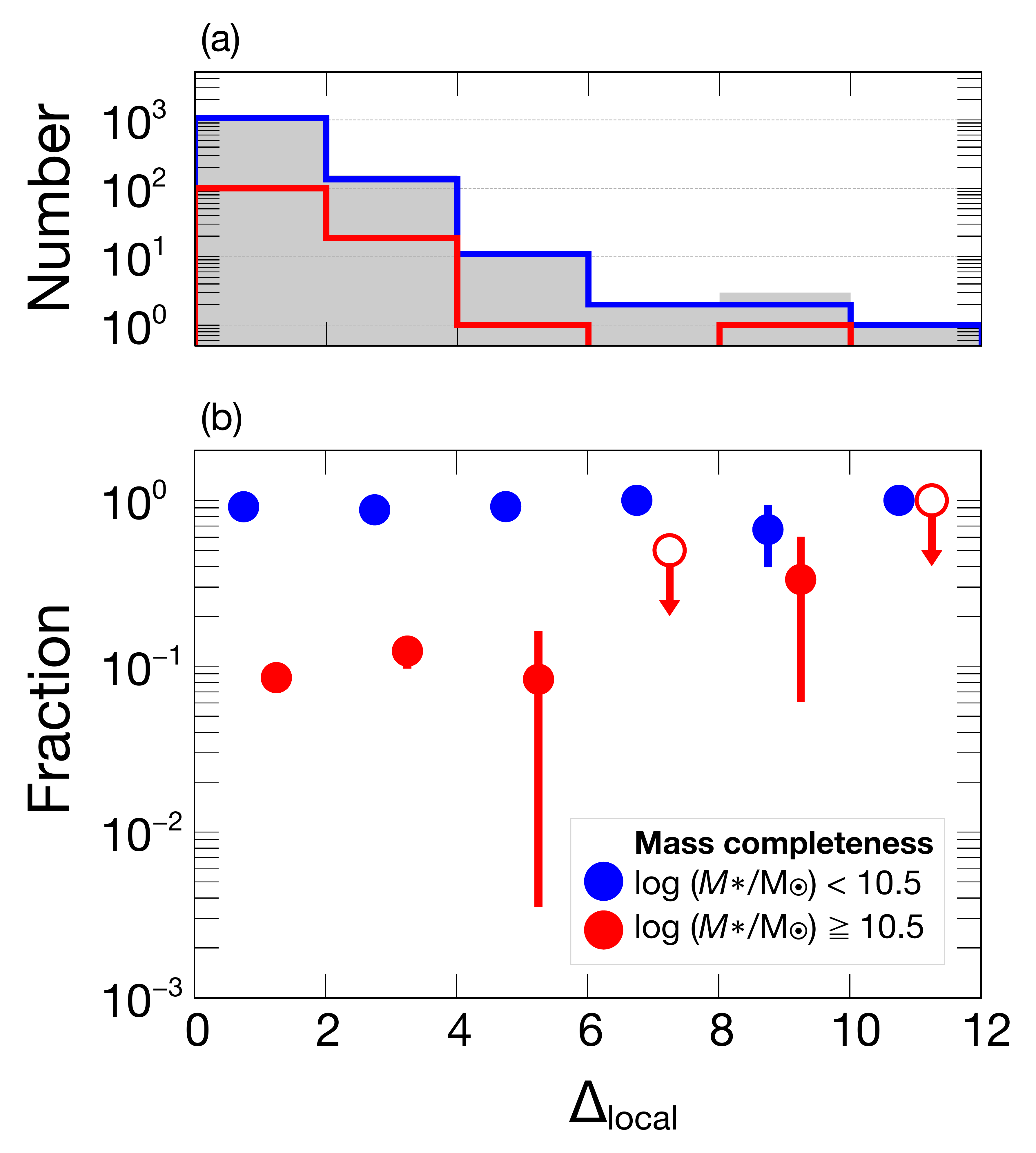}
\caption{
(a) : Number histograms of galaxy samples as as function of the density contrast. 
(b): Fractions of low- and high-mass galaxies as a function of the density contrast.
For both panels in (a) and (b), blue and red lines or markers represent low- and high-mass galaxy samples respectively. The upper limit of high-mass galaxies is also shown as white circles when none is found in a bin. 
We consider an estimate to be the upper limit, assuming one sample in a bin ($N_\text{bin}=1$).
These measurements are performed in each $\Delta_\text{local}=2.0$ bin. 
The errors are calculated through a Poissonian approach as $\sqrt{n}$/\textit{N}, where \textit{n} and \textit{N} are the numbers of targeting sample and total galaxies within an each $\Delta$ bin.
}
\label{fig:frac_MS_dDM}
\end{figure}

We demonstrate the actual number of galaxy samples to a certain density contrast in Figure~\ref{fig:frac_MS_dDM}-a. More than $80\%$ of our samples are in underdense regions in $0\leq\Delta_\text{local}<2$. 
We also show the fraction of low-/high-mass galaxies at a given density contrast $\Delta_\text{local}$ in Figure~\ref{fig:frac_MS_dDM}-b. Qualitatively, the fraction of low-mass galaxies is always dominant over the full $\Delta_\text{local}$ range, whereas the fraction of high-mass galaxies gradually increases by about $20-30~\%$ increasing the density contrast.
However, those differences in the fraction of low- and high-mass galaxies as a function of the density contrast were statistically insignificant in the two-sample Kolmogorov–Smirnov test with $p$-value $=0.18$. 
We note that the results on low-mass galaxies would not be largely changed if we exclude galaxies with $\Mstar<10^9$~M$_\odot$ since its fraction is small enough ($3\%$).

\subsubsection{Star Formation Activities}
\label{sec:results_delta_physical}

%
\begin{figure}[t]
\includegraphics[width=\linewidth]{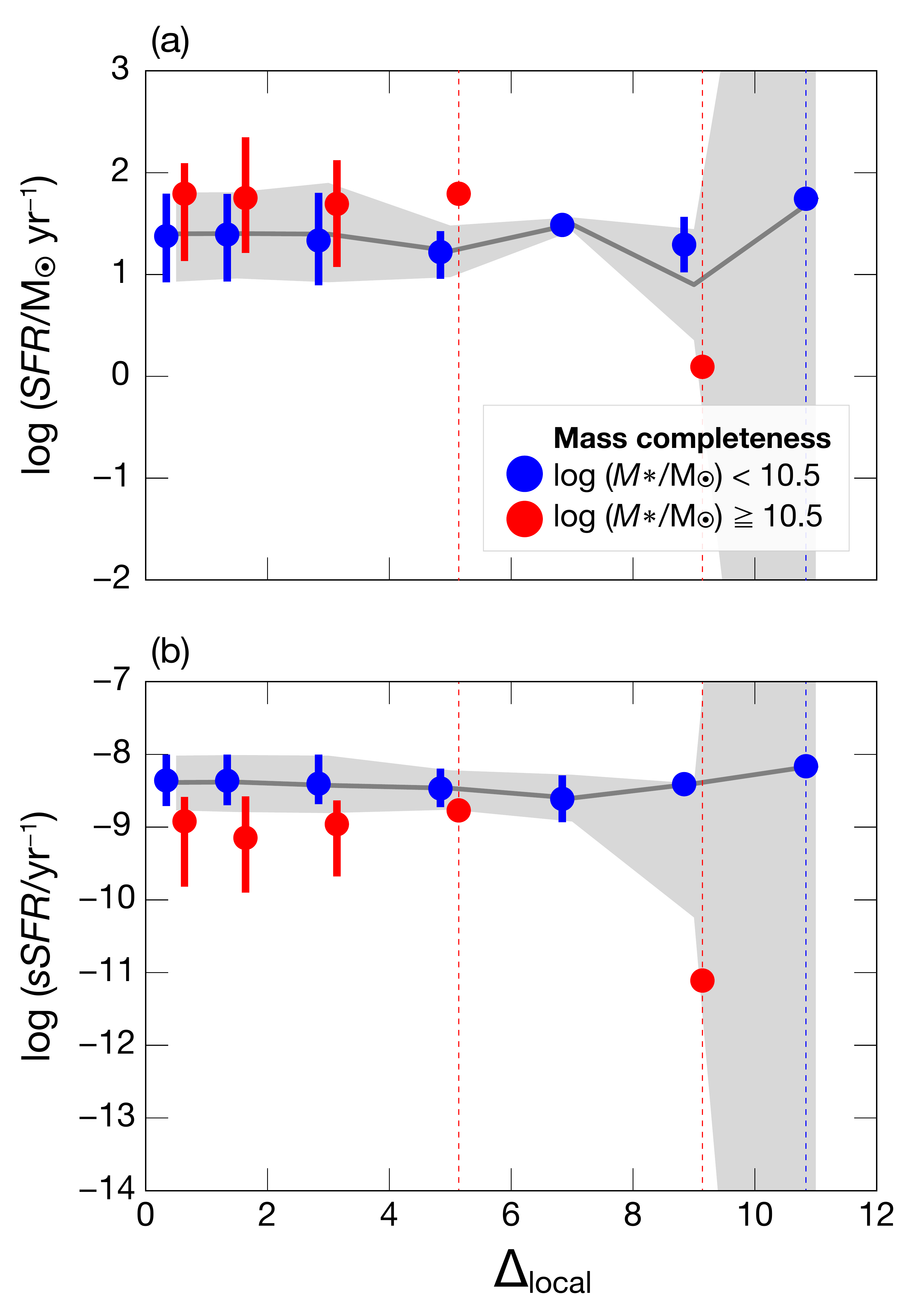}
\caption{
The median (a) \textit{SFR} and (b) \textit{sSFR} of low-/high-mass galaxies as a function of the density contrast. They are plotted in each 
$\Delta_\text{local}=1.0$ at $\Delta_\text{local}<2.0$ but each $\Delta_\text{local}=2.0$ at $\Delta_\text{local}\geq2.0$.
The error bars indicate the percentiles 16th and 84th of \textit{SFR} and \textit{sSFR}. A gray line and shades are the medians and 68 percentiles of the entire mass-completeness sample. 
In the case of only one sample in a bin, we add an infinite error by dashed lines or grey shades because we cannot evaluate the 16th and 84th percentiles.
}
\label{fig:SFA_dDM}
\end{figure}

Figure~\ref{fig:SFA_dDM} represents the median \textit{SFR} and \textit{sSFR} as a function of density contrast. 
We should note that this study is the first to investigate galaxy trends in underdensities ($\Delta_\text{local}<1$) beyond $z\gtrsim 1$. 

The environmental trends for \textit{SFR} is seen in Figure~\ref{fig:SFA_dDM}-a. 
We first find in low-mass galaxies that \textit{SFR} is nearly constant about at $\log$(\textit{SFR}) $=1.3-1.4$ over the full $\Delta_\text{local}$ range. Note that although it has a factor of $4$ difference between their lowest and highest values of \textit{SFR}, it is still within the errors. 
The \textit{SFR} of high-mass galaxies is also independent of the density contrast until $\Delta_\text{local}=6$ within the error. In addition, we find that high-mass galaxies have possibly higher \textit{SFR} by a factor of $2.5-3.1$ than low-mass galaxies within this $\Delta_\text{local}$ range, though it is within an allowable error range. This implies that the \textit{SFR} of both low- and high-mass galaxies do not depend on the density contrast up to some density level. 
Just as with the trend of \textit{SFR}, the \textit{sSFR} shows little change over the wide $\Delta_\text{local}$ range in low-mass galaxies and at $\Delta_\text{local}<6$ in high-mass galaxies (see in Figure~\ref{fig:SFA_dDM}-b). 

Both \textit{SFR} and \textit{sSFR} exhibit a rapid decline in median values on high-mass galaxies beyond $\Delta_\text{local}=6$. Nonetheless, that may not be statistically significant due to substantial uncertainty caused by the small sample size ($N_\text{bin}=1$). We will discuss more about it in Section~\ref{sec:dis_galevo_SFA}.

\subsubsection{
Mergers, SMGs, AGNs, and QGs 
}
\label{sec:results_delta_type}

%
\begin{figure*}[ht]
\includegraphics[width=0.9\linewidth]{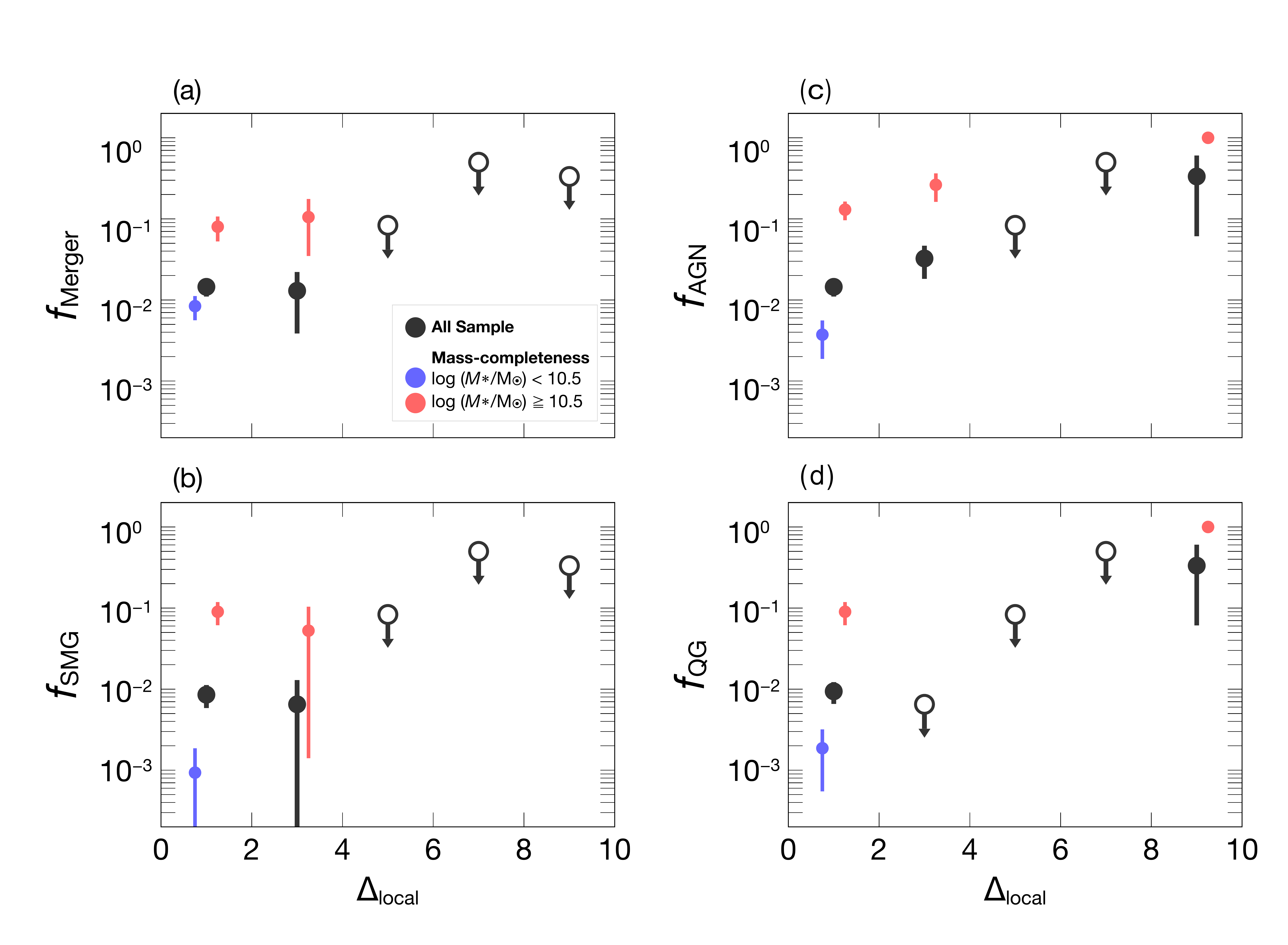}
\caption{
The fraction of different galaxy types in each $\Delta_\text{local}=2.0$ bin. Each sub panel demonstrates individual special galaxy type fractions of (a) mergers, (b) SMGs, (c) AGNs, and (d) QGs. Black, blue, and red circles represent fractions estimated from the mass-complete (i.e., all galaxies), low-, and high-mass galaxy samples.   
The errors are calculated through a Poissonian approach as $\sqrt{n}$/\textit{N}, where \textit{n} and \textit{N} are the numbers of targeting galaxy type and total galaxies within an each $\Delta_\text{local}$ bin.
The upper limit is also shown as white circles when no galaxy sample is found in a bin. 
We consider an estimate to be the upper limit, assuming one sample in a bin ($N_\text{bin}=1$).
Data points are slightly shifted horizontally for clarity.
}
\label{fig:frac_type_dDM}
\end{figure*}

Figure~\ref{fig:frac_type_dDM} demonstrates the fractions of galaxies that are mergers, SMGs, AGNs, and QGs as a function of the density contrast. We should caution that these samples are small due to limited availability of spec-$z$ measurements on these classes. We have 19 mergers, 11 SMGs, 23 AGNs, and 12 QGs. Dividing these galaxy types into different masses yield even smaller subsamples. Details of their sample sizes are provided in Table~\ref{tab:galaxy_type}.

Here we present four main results. 
First, we find an absence of mergers and SMGs in higher-density regions, particularly at $\Delta_\text{local}\geq4$.
Interestingly, this $\Delta_\text{local}$ range for the presence of mergers and SMGs, i.e., $\Delta_\text{local}=0-4$, overlaps the $\Delta_\text{local}$ range, which shows a lack of dependence of \textit{SFR} and \textit{sSFR} on the density contrast regardless of mass bin (see Section~\ref{sec:results_delta_physical}).

The second trend is the independence of mergers and SMGs fractions from the density contrast when we limit to the range where samples are found. The lack of fractional changes on the density is also apparent in the high-mass regime of both mergers and SMGs because more than half of those samples belong to the high-mass regime. 

The third is the presence of AGNs and QGs in high-density environments at $\Delta_\text{local}\geq4$.
In addition, quite intriguingly, nearly all high-mass galaxies found in high density are either AGNs or QGs by showing $f_\text{AGN}$ or $f_\text{QG}\sim1$.

The fourth is the increase of AGNs and QGs fractions as a function of density contrast. The same increases are accentuated when the sample is limited to high-mass galaxies. 

There are no onservational works which we can directly compare our results for now because all past investigations in the fractions of mergers, QGs, SMGs, and AGN have been done based on the \textit{galaxy} overdensity. Nonetheless, the comparison across studies would be helpful to obtain the deep insight into our results. 

One of the frequently studied galaxy fraction is QG fraction. 
Likewise our results, some studies at lower redshifts have observed the increase of QG fractions with galaxy overdensities (e.g., \cite{Chuter11,Quadri12,Kawinwanichakij17,Reeves21}).
\citet{Kawinwanichakij17} have evaluated the QG fraction at $0.5<z<2.0$ on the galaxy overdensity and observed the increase of the fraction. 
\citet{Reeves21} have shown that QG fractions in $\Mstar>10^{11}$~M$_\odot$ at $1<z<1.5$ are systematically larger in their group samples than the field. 
In contrast, the opposite trend, i.e., an anti-correlation of QG fraction and galaxy overdensity, has been found in some studies (e.g., \cite{Tran10,Lin16}). 
A lack of dependence on galaxy overdensity has also been reported (e.g., \cite{Darvish16}). Therefore, a solid conclusion in QG fraction can not yet be arrived at, although our study is to our knowledge the first explicit analysis of environmental properties of QGs at $z>2$.

The AGN fraction is also often examined in the literature. Many studies have suggested that AGN fraction depends on the galaxy environment at low-redshifts, in which the AGN fraction is found to be higher in underdense galaxy environments than overdense ones (e.g., \cite{Kauffmann04,Silverman09,Linden10,Lopes17}). We should note that some studies have argued for no significant preference on galaxy overdensity for AGN fraction at the same low-redshift range (e.g., \cite{Miller03,Man19}). 
In contrast, \citet{Lehmer09} have detected a factor of $6.1$ larger AGN fraction in the SSA22 protocluster at $z=3.09$ than in the field. Similarly, the enhancements of AGN fractions have also been reported in protocluster or cluster environments at $z>2$ (\cite{Digby-North10,Polletta21}).
This positive correlation between the AGN fraction and galaxy overdensities is clearly opposite to the one seen in low-$z$ universe and could be more significant beyond $z=2$ (\cite{Martini13,Krishnan17}). Our results on the AGN fractions and its increase with density contrast $\Delta_\text{local}$, would agree with this positive correlation found at higher redshift.
However, a lack of AGN enhancement in an individual protocluster at $z\sim2$ has been observed by \citet{Macuga19}, suggesting that there is still room for more discussion on the relation between the AGN fraction and the environment. Further work will be required to conclude that relationship. 

Unlike AGN and QG fractions, the density dependence of mergers and SMG activity has not been well studied in the literature.
The merger fraction in the $z>2$ universe has generally been inconclusive. 
Some studies have observed a higher merger fraction in clusters than in fields (e.g., \cite{Lotz13,Hine16,Watson19}), while other studies have found no enhancement in clusters (e.g., \cite{Delahaye17}). The lack of enhancement or even the absence of mergers in the higher density regions in our data may be indicative of agreement with the latter studies or simply a result of the small sample size. We will discuss it in Sections~\ref{sec:dis_impact_type} and \ref{sec:dis_galevo_type}. 
Many observations have reported that dusty star-forming galaxies represented by SMGs are often found in massive halos (e.g., \cite{Venemans07,Hickox12,Umehata15,Wilkinson17,Jones17,Zeballos18, Crespo21}). In contrast, \citet{Miller15} have claimed that SMGs at $z\leq2.5$ are supposed to reside in less dense environments, not the most massive overdensities from their semi-analytic simulations. 
Some observations have also drawn similar conclusions (e.g., \cite{Chapman09,Casey16}). Hence, the true relation between SMG prevalence and overdensity is still unclear.

\subsection{Dependence On The Stellar Mass}
\label{sec:result_mass}

%
\begin{figure}[ht]
\includegraphics[width=\linewidth]{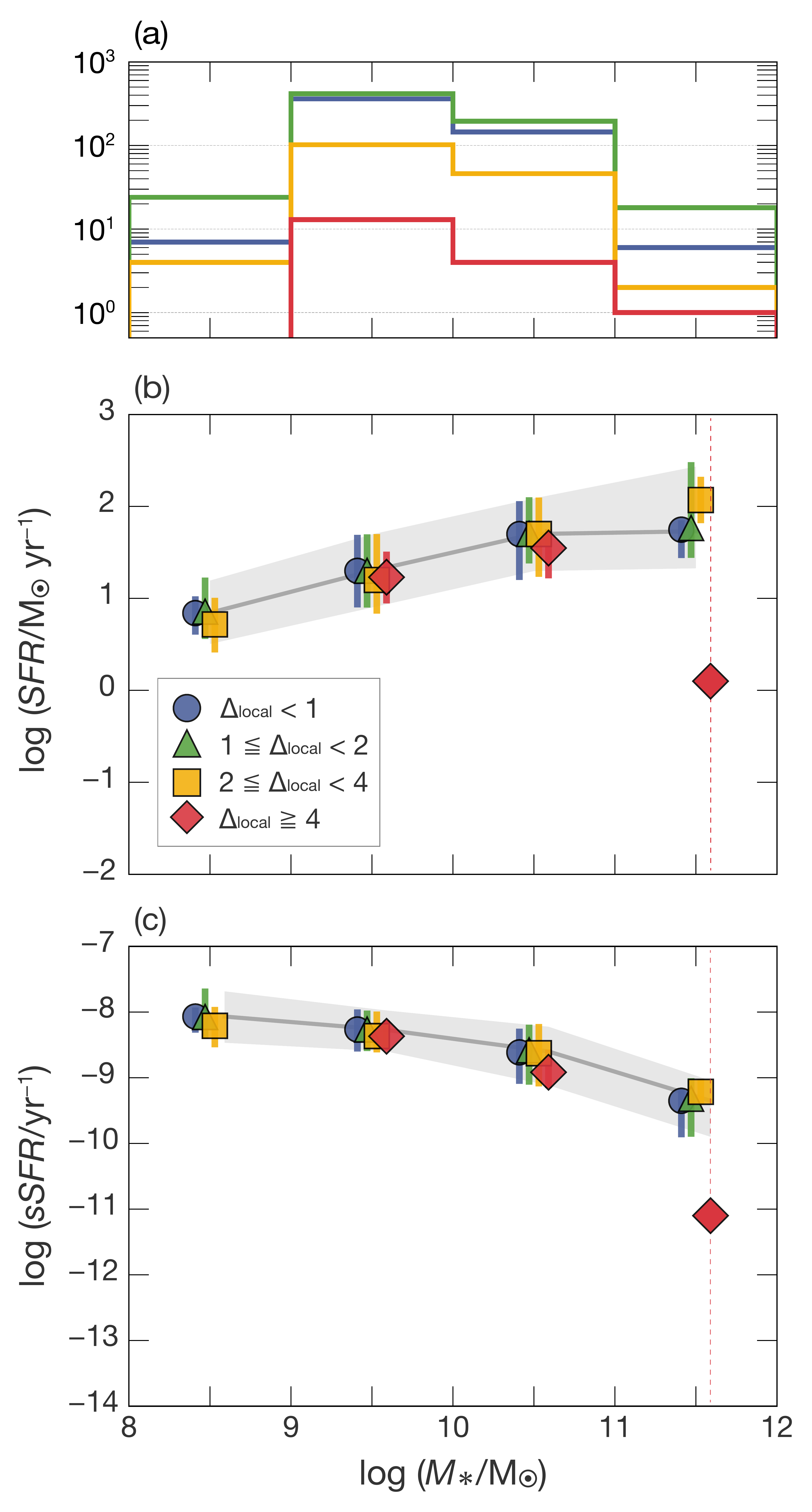}
\caption{
(a) Number histograms of matter density subsamples in each $\Mstar=10^1$~M$_\odot$ bin from $\Mstar=10^8$ to $10^{12}$~M$_\odot$. 
The median (b) \textit{SFR} and (c) \textit{sSFR} to $\Mstar$ in each $\Mstar=10^1$~M$_\odot$ bin. 
For all panels, blue, green, yellow, and red indicate subsamples in $\Delta<1$, $1\leq\Delta<2$, $2\leq\Delta<4$, and $\Delta\geq4$. 
The error bars indicate the percentiles 16th and 84th of \textit{SFR} and \textit{sSFR}. A gray line and shades are the medians and 68 percentiles of \textit{SFR} and \textit{sSFR} of the entire mass-completeness sample. Data points are slightly shifted horizontally for clarity.
We give an infinite error bar to a data point at $\Mstar\geq10^{10}$~M$_\odot$ in $\Delta\geq4$ subsample because there is only one sample in that bin.
}
\label{fig:SFA_Ms_dDM}
\end{figure}

Galactic properties are shaped and influenced by not only their environments but also their internal processes. We also examine the impact of the internal process on galactic properties and galaxy types in this subsection.

\subsubsection{Star Formation Activities}
\label{sec:result_massSFA}

We plot the star formation and specific star-formation rates as a function of stellar mass for galaxies at different density contrasts in Figure~\ref{fig:SFA_Ms_dDM}. 
Overall, all density subsamples follow the so-called star formation main sequence as is reported in the literature (e.g., \cite{Daddi07,Kashino13,Whitaker14,Speagle14,Tomczak16}). Meanwhile, the opposite (anti-correlation) is seen in the \textit{sSFR} vs $\Mstar$ relation.
The median \textit{SFR} and \textit{sSFR} are generally independent of the underlying matter density. A possible exception is at $\Mstar\geq10^{11}$~M$_\odot$, when the high-$\Delta$ subsample ($\Delta_\text{local}\geq4$) shows drops in \textit{SFR} and \textit{sSFR}. However, a single data point in this bin can preclude a firm conclusion from being reached. We test it in more detail in Section~\ref{sec:dis_galevo_SFA}.

\subsubsection{Mergers, SMGs, AGNs, and QGs}
\label{sec:result_massType}

%
\begin{figure}[t]
\includegraphics[width=\linewidth]{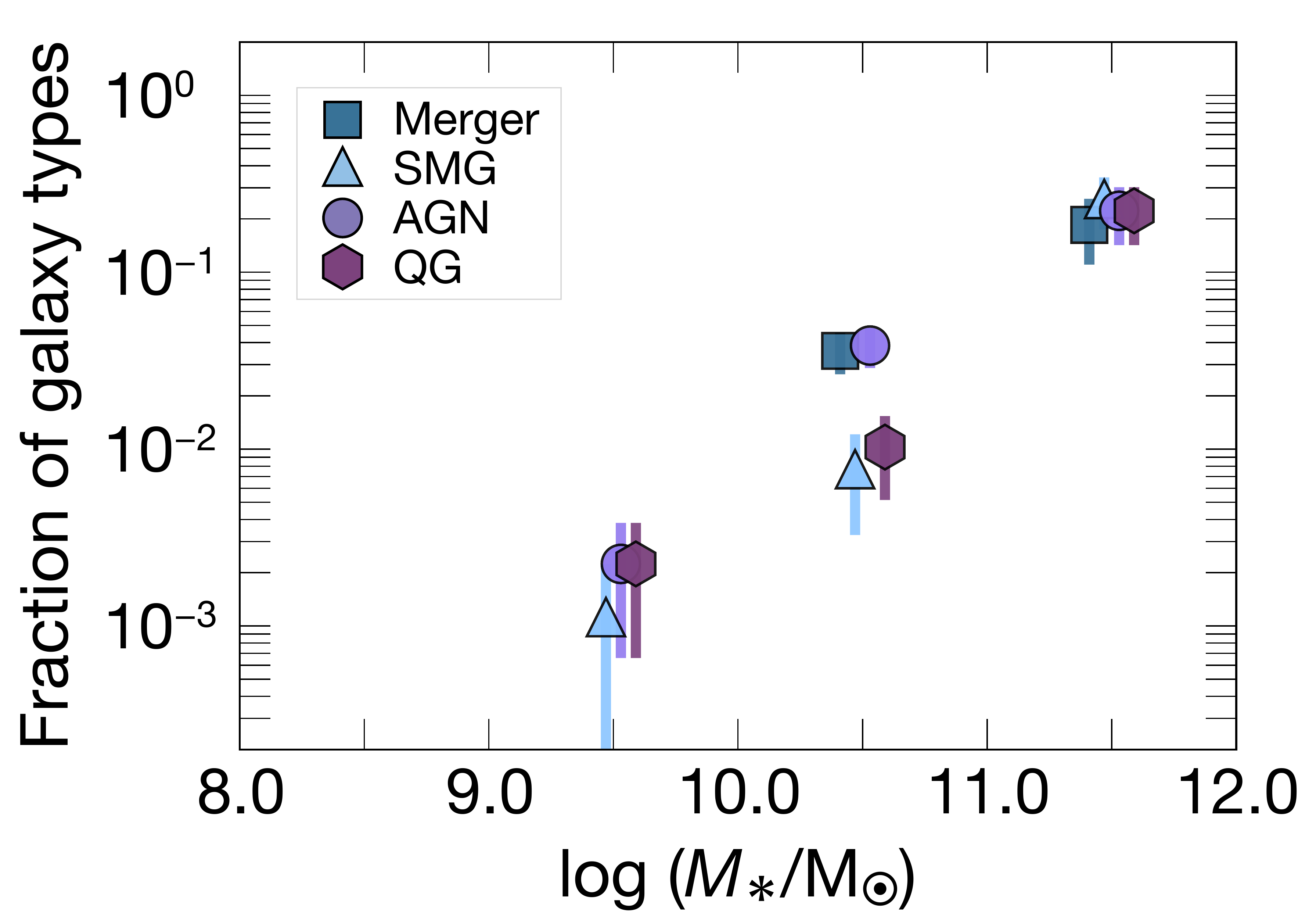}
\caption{
The fraction of different special galaxy types in different logarithmic $\Mstar$ bins from $10^8$ to $10^{12}$~M$_\odot$ obtained from all mass-complete samples in the COSMOS field at $2<z<2.5$. Different markers and colors represent individual special galaxy types -- a square, triangle, circle, and hexagon are mergers, SMGs, AGNs, and QGs. 
Data points are slightly shifted horizontally for clarity. Vertical lines indicate boundaries for each $10^{1}$~M$_\odot$ bin.
}
\label{fig:frac_type_Ms}
\end{figure}

Figure~\ref{fig:frac_type_Ms} demonstrates the fraction of each galaxy type as a function of $\Mstar$. 
The notable feature of the plot is a monotonic increase in the fraction with increasing $\Mstar$ for all galaxy types. 
Unlike Figure~\ref{fig:frac_type_dDM}, mergers and SMGs are also found across a wide $\Mstar$ range, with increasing fractions.

The increase of these galaxy types in the mass has also been reported in the literature --- 
mergers (e.g., \cite{Bundy09}),
SMGs (e.g., \cite{Gonzalez11}),
AGNs (e.g., general trend \cite{Brusa09,Tanaka12,Pimbblet13,Lopes17}, 
for radio-mode AGN \cite{Sabater19,Miraghaei20}), and 
QGs (e.g., \cite{Sobral11,Darvish16,Kawinwanichakij17,Reeves21}). 
Those studies, together with ours in Figure~\ref{fig:frac_type_Ms} would support the galaxy evolution driven by mass. We will discuss it in more detail in Section~\ref{sec:dis_galevo_type}.

\section{Discussion}
\label{sec:dis}

\subsection{The Impact Of Incomplete Samples To Our Results}
\label{sec:dis_impact_type}

Quantifying the impact of a biased sample on the results seen in the mergers, AGNs, SMGs, and QGs is challenging because our spectroscopic sample are only a small fraction of their original catalogs. For instance, our merger and SMG samples are equivalent to only $24-30\%$ and $5-8\%$ of the original samples.\footnote{We only count galaxies whose photo-$z$ are $1.9\leq z\leq2.6$ within the respective merger, SMG, AGN, or QG catalogs as original samples.}
This small sample size might have enhanced the random chance of selecting galaxies only at lower matter density regions, even in the absence of a physical mechanism. Likewise, the AGNs and QGs within our maps comprise $3-4\%$ and $0.8\%$ of the full AGN and QG samples, respectively. The small number statistics could potentially cause a spurious correlation between these object fractions as a function of the density contrast, due to statistical fluctuations. 
We, therefore, briefly estimate the chance probability of detecting a correlation between the respective population fractions and the density contrast, assuming that $\Delta_\text{local}$ around galaxies is randomly determined. 
We randomly distributed mock sources with the same sample size as the mass-complete sample across the COSTCO density field and assessed the mock fraction-$\Delta_\text{local}$ relation, i.e., the same plot of Figure~\ref{fig:frac_type_dDM}, by selecting the same number of mock sources with the same total number of original galaxy type shown in Table~\ref{tab:galaxy_type}. This test was repeated $1,000$ times to evaluate the chance probability.

First, we calculate the probability for finding no mergers and SMGs in the higher density regions due to the small sample sizes.
The random probabilities for a spurious absence in high-density regions is found to be less than $2\%$ for all mock merger and SMG samples. For those reasons, we believe that the lack of mergers and SMGs at higher density regions is unlikely to be due to random chance. 

Second, we evaluate the probability that the increasing AGN and QG fractions as a function of the density contrast could have occurred from pure chance. We obtain high probabilities of greater than $66\%$ for mock AGNs and QGs to exhibit a correlation with density contrast.
This is because in the real data, the number of these objects in each $\Delta$ bin becomes smaller with $\Delta_\text{local}$ as seen in Figure~\ref{fig:frac_MS_dDM}, whereas the mock AGNs and QGs are randomly distributed across the $\Delta$ bins. Thus, we cannot rule out the possibility of their random distributions on the density contrast.
Although we will conduct the following discussions based on the obtained fractional results, solid conclusions would be made from unbiased, complete data taken by future telescopes (e.g., Giant Magellan Telescope, Thirty Meter Telescope, and Extremely Large Telescope) and instruments.

\subsection{The Mechanism For Galaxy Evolution}
\label{sec:dis_galevo}

In Section~\ref{sec:results}, we studied the relation between galactic properties (i.e., star formation activities and galaxy types) and either underlying matter density contrast or stellar mass. Here we recap those results and discuss the possible mechanisms for galaxy evolution and subsequent star formation quenching.

\subsubsection{The Implication from Star Formation Activities}
\label{sec:dis_galevo_SFA}

%
\begin{figure}[t]
\includegraphics[width=\linewidth]{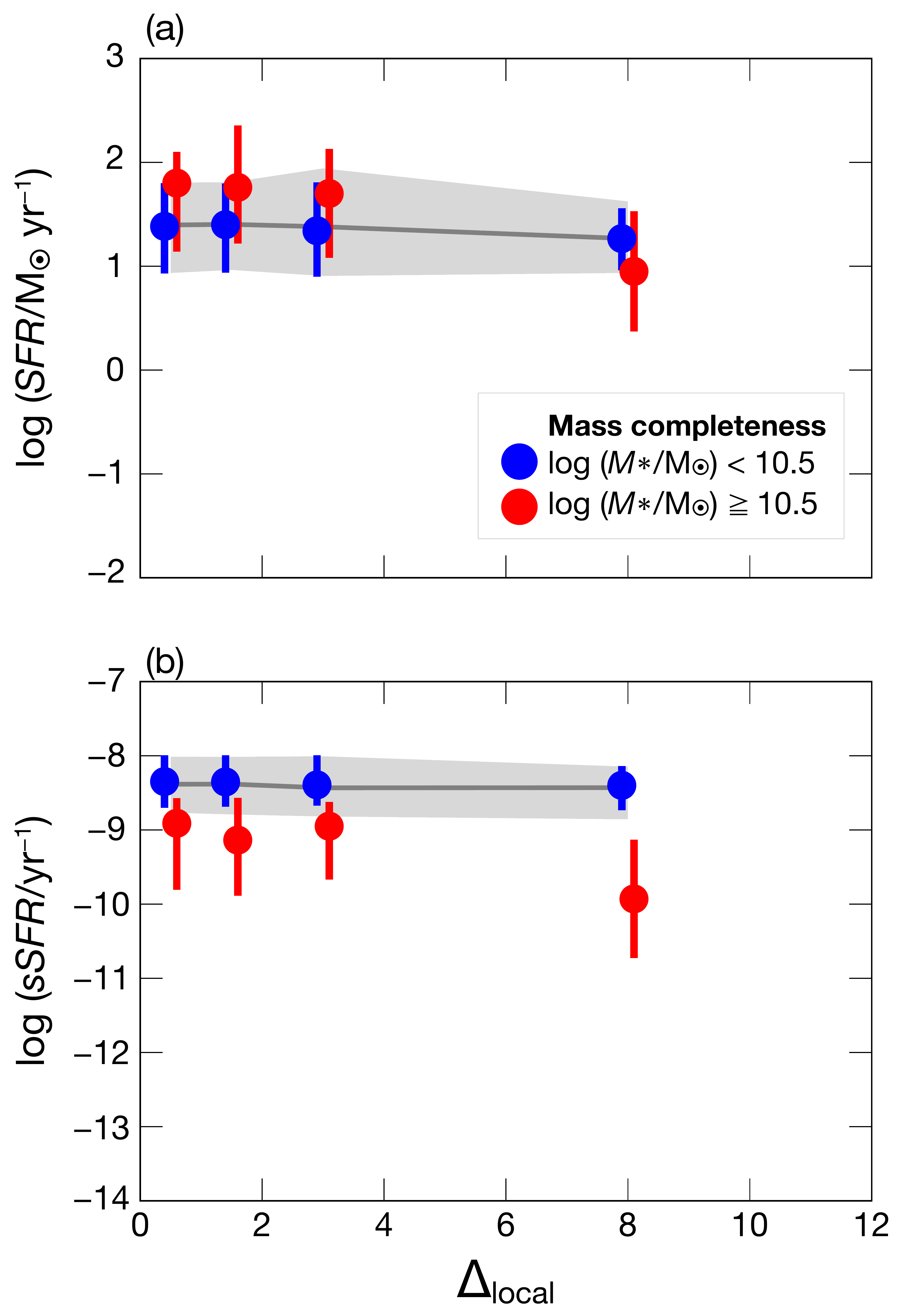}
\caption{
The median (a) \textit{SFR} and (b) \textit{sSFR} of low-/high-mass galaxies as a function of matter overdensity. The $\Delta_\text{local}$ bin size is [1, 1, 2, 8] to increase the sample size in the final bin. 
The 16th and 84th percentiles are used as errorbars.
}
\label{fig:SFA_dDM_rebin1}
\end{figure}

\begin{figure}[t]
\includegraphics[width=\linewidth]{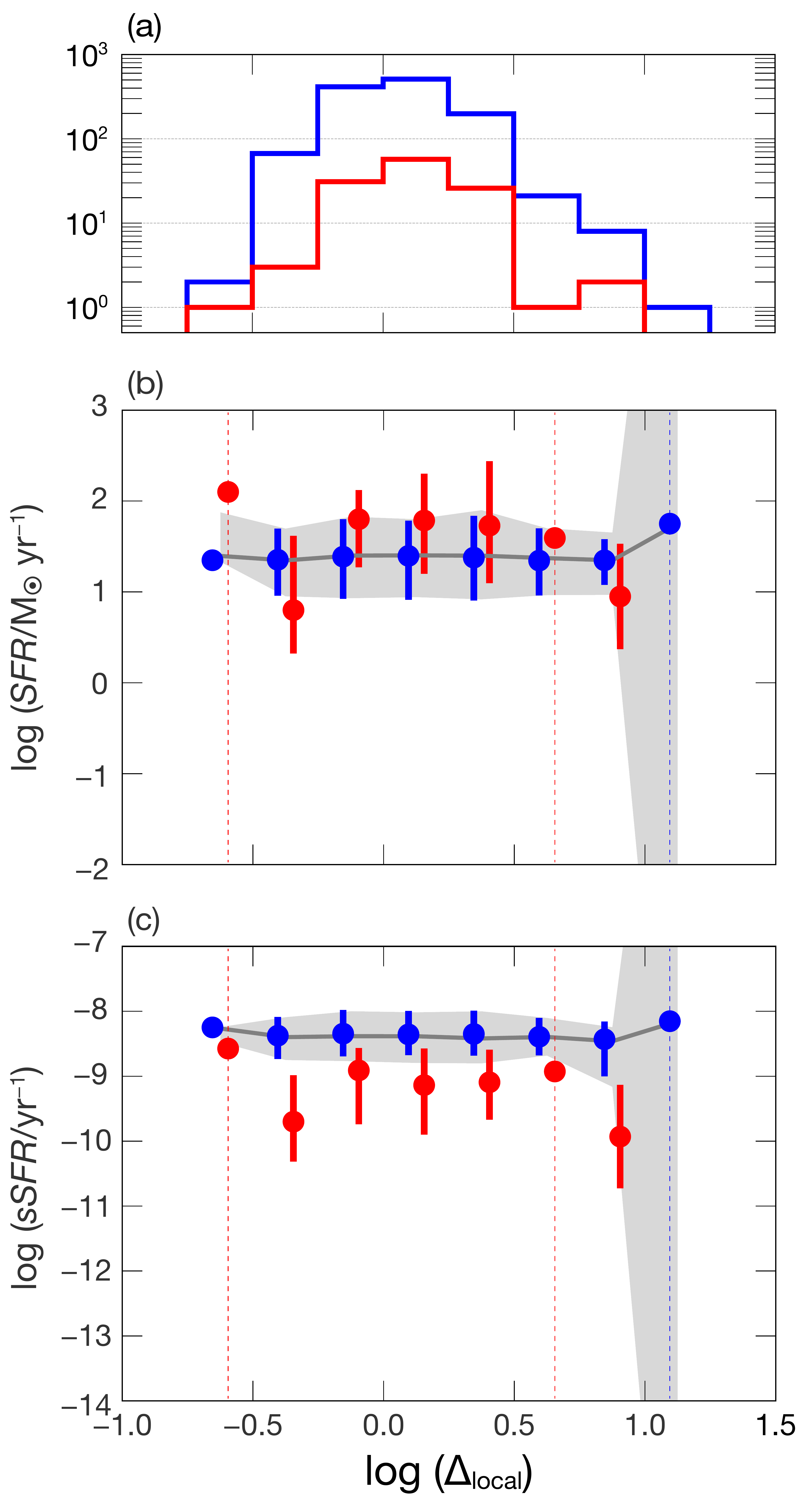}
\caption{
(a) Number histograms of matter density subsamples in each $\log (\Delta_\text{local})=0.25$ bin from $\log (\Delta_\text{local})=-1.0$ to $1.5$.
The median (b) \textit{SFR} and (c) \textit{sSFR} of low-/high-mass galaxies as a function of matter overdensity, in log scale. The bin size is $\log (\Delta_\text{local})=0.25$. 
The 16th and 84th percentiles are used as errorbars.
}
\label{fig:SFA_dDM_rebin2}
\end{figure}

\begin{figure}[t]
\includegraphics[width=\linewidth]{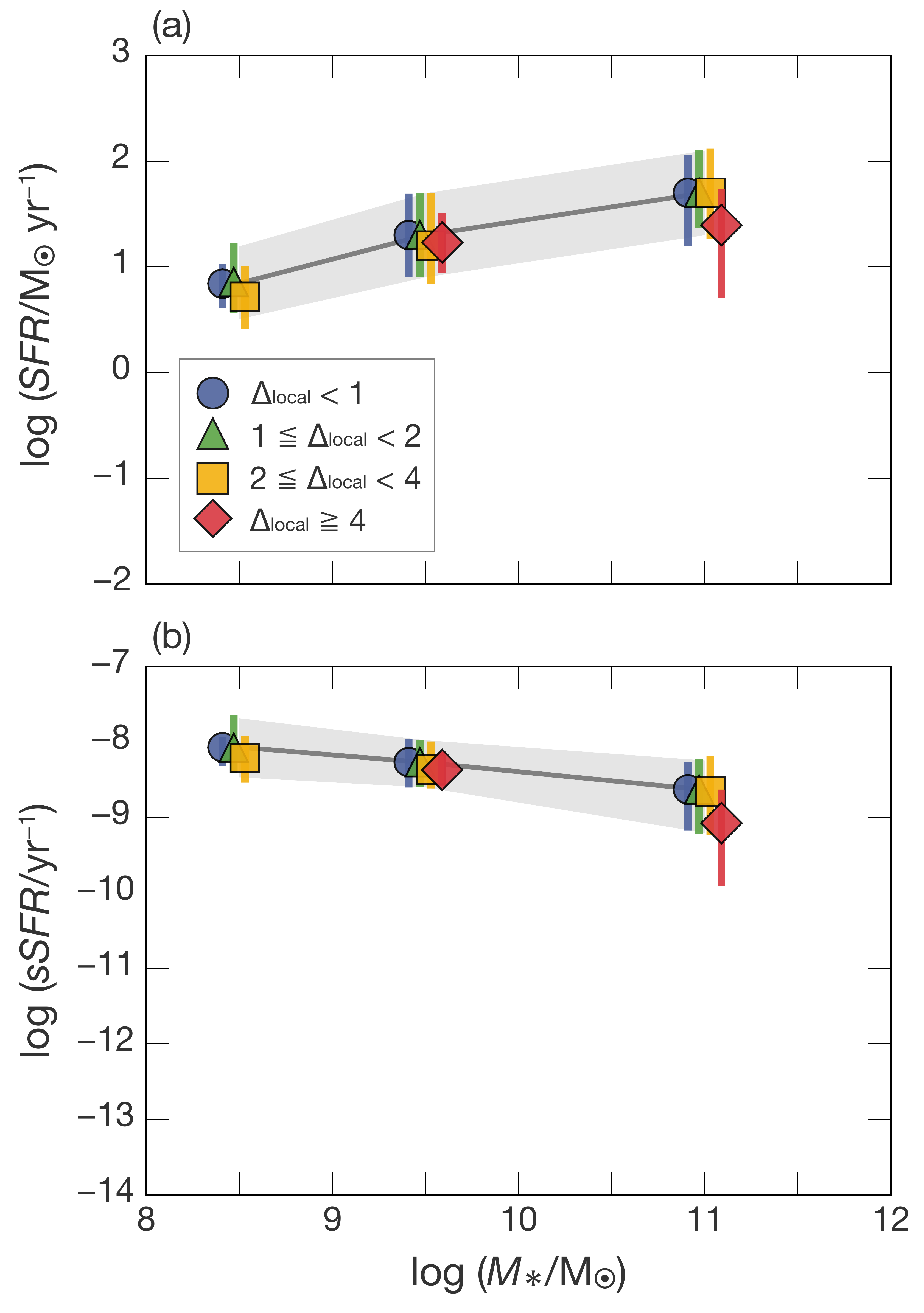}
\caption{
The median (a) \textit{SFR} and (b) \textit{sSFR} as a function of stellar mass, in log scale, for matter density subsamples. The $\log{\Mstar}$ bin sizes are [1, 1, 2]. The usage of symbols and lines are the same as Figure~\ref{fig:SFA_Ms_dDM}.
}
\label{fig:SFA_Ms_rebin1}
\end{figure}

In Section~\ref{sec:results_delta_physical}, we found no dependence of both \textit{SFR} and \textit{sSFR} on the density contrast for low-mass galaxies ($\Mstar<10^{11}$~M$_\odot$). That is also confirmed in the main sequence diagram in Figure~\ref{fig:SFA_dDM}, in the sense of the lack of dependence among the density subsamples.
High-mass galaxies also showed that star formation activities do not depend on the density contrast within the uncertainties. In contrast, their median values demonstrated a decline in high-density regions beyond $\Delta_\text{local}=4-6$, as seen in Figures~\ref{fig:SFA_dDM} and \ref{fig:SFA_Ms_dDM}. The significance of these declines, however, is low because the median values on \textit{SFR} and \textit{sSFR} are from only one sample in the bin showing a decline.  
We more thoroughly evaluate the significance of the declines in star formation activities for high-mass galaxies in high-density regions by re-binning $\Delta_\text{local}$ and $\Mstar$.

We apply the re-binning for $\Delta_\text{local}$ in two ways. The first is to combine all $\Delta_\text{local}$-bins greater than $4$ into one to boost the sample size for high-mass galaxies in high-density regions, as presented in Figure~\ref{fig:SFA_dDM_rebin1}. The other way is to evaluate the star formation activities on $\Delta_\text{local}$ in log-scale, as shown in Figure~\ref{fig:SFA_dDM_rebin2}. 
In both cases, \textit{SFR} and \textit{sSFR} for high-mass subsamples do not significantly vary on the $\Delta_\text{local}$ within the error, albeit their median values decline by an order of magnitude at the densest bin of $\Delta_\text{local}\geq4$.

For $\Mstar$, we re-bin by combining all bins at $\Mstar\geq10^{10}$~M$_\odot$ into one to increase the sample size for the high-$\Delta$ subsample. Figure~\ref{fig:SFA_Ms_rebin1} demonstrates the results. The overall trend is the same as seen in Section~\ref{sec:result_massSFA} that both \textit{SFR} and \textit{sSFR} values are comparable in each $\Mstar$ bin, regardless of $\Delta$ subsamples. Although their median values for high-$\Delta$ subsample show a small drop by about $0.5$ dex at $\Mstar\geq10^{10}$~M$_\odot$, they are still within the error.

To conclude these tests via re-binning, we find comparable in \textit{SFR} and \textit{sSFR} within the uncertainties, regardless of stellar mass or $\Delta$ of galaxies. In other words, star formation activities in galaxies at $z=2$ are generally independent of the local matter density but depend on the stellar mass. This implies that the stellar mass of galaxies plays a major role in star formation activities and quenching. 
This is consistent with some observational studies based on \textit{galaxy} overdensities. According to \citet{Darvish16}, the median \textit{SFR}s of galaxies at $z>1$ are almost independent of their overdensities. In addition, they have also found the independence of the median \textit{SFR} on the overdensity at fixed stellar mass bins for star-forming galaxies out to $z\sim3$. 
\citet{Chang22} have studied green valley galaxies and observed no significant difference in \textit{sSFR} between the highest and lowest density environments at $1.0<z<2.5$.

Despite these results, it is noteworthy that we still identify the decline of the median of \textit{SFR} and \textit{sSFR} for high-mass galaxies inside overdense regions in all tests, as seen in Sections~\ref{sec:results_delta_physical} and \ref{sec:result_massSFA}. 
A similar dependence of star formation activities on \textit{galaxy} overdensities has been reported in the literature. 
\citet{Chartab20} have found the anti-correlation of the \textit{SFR} with overdensities for their massive samples in $\Mstar\geq10^{11}$~M$_\odot$ at $z<3.5$; however they have not detected its significant environmental dependence for their low mass samples ($\Mstar<10^{11}$~M$_\odot$).
\citet{Lemaux22} have observed a decline of \textit{SFR} with galaxy overdensities for galaxies in the densest regions at $2.0<z<2.7$. 
If these phenomena hold true, it may suggest the additional environmental effects or other processes working for galaxies satisfying some conditions (i.e., high mass and within overdensities).
The ``overconsumption'' of cold gas could be a viable physical process for such circumstances, which occur only in high-mass galaxies in higher-density regions. 
The timescale of gas depletion ($\tau_\text{dept}$) for forming stars is proportional to the ratio of gas mass to \textit{SFR} ($\tau_\text{dept} \propto M_\text{gas}$/\textit{SFR}). 
Since \textit{SFR} depends on both stellar mass and redshift, $\tau_\text{dept}$ gets shorter in high-mass galaxies than low-mass ones in the case of the same amount of gas. 
On the other hand, $M_\text{gas}$ can be determined by the balance of secular outflows and inflows. Cosmological hydrodynamic simulations by \citet{vandeVoort17} have suggested that the gas accretion rate is suppressed in denser environments, which leads to a shorter gas depletion of galaxies in denser environments. This more significantly affects high-mass galaxies at denser environments and results in the suppression of their star formation on short timescales (see also \cite{Balogh16,Kawinwanichakij17,Chartab20}), known as the overconsumption. 
Our high-mass samples in high-density regions might also suffer from this overconsumption of gas reservoirs and thus show sudden drops in their star formation activity. However, again, we cannot conclude this scenario from our current results due to the significant uncertainties.

\subsubsection{The Implication from Merger, SMG, AGN, and QG Fractions}
\label{sec:dis_galevo_type}

The spectroscopic samples of mergers, SMGs, AGN, and QGs comprise only a small fraction of the main galaxy samples, as discussed in Section~\ref{sec:dis_impact_type}. This brings large uncertainties in our results shown in Section~\ref{sec:results_delta_type}. Nonetheless, it is worth providing some possible scenarios for star formation quenching and environmental dependence based on our results.

A strong dependence on the merger, SMG, AGN, and QG fractions is found on the stellar mass, $\Mstar$. All four categories show a positive correlation of their prevalence with $\Mstar$ (Figure~\ref{fig:frac_type_Ms}). 
That would be another support that mass is a main driver of galaxy quenching.

By contrast, the density dependence is different between these categories. Galaxies show increasing AGN and QG fractions with their underlying density.  
A higher QG fraction at higher density regions could be reasonable, given the lower star formation activities of high-mass galaxies seen at $\Delta_\text{local}\geq4-6$. 
The increase of AGN and QG fractions with the density contrast can also be explained by the linear theory prediction and Figure~\ref{fig:frac_type_Ms}. In other words, massive galaxies that have a greater probability of hosting AGNs and/or QGs tend to be located in high-density regions, resulting in a positive correlation with the local density. 
In contrast, the remaining two types, mergers and SMGs show no clear dependence on the density and are even devoid at higher density regions in $\Delta_\text{local}\geq4$. 
Those results contradict the scenario predicted by linear theory and Figure~\ref{fig:frac_type_Ms}, in which all these populations become more correlated with $\Mstar$ and therefore with the density as well. 

We discuss some possible causes of the contradiction for no mergers and SMGs within our spec-$z$ samples in higher-density regions. Firstly, that could be simply due to random chance. However, this situation seems to be unlikely, as discussed in Section~\ref{sec:dis_impact_type}. 
Secondly, it results from the coevolution of galaxies and their surrounding cosmic web. 
In other words, galaxies fall into higher density regions while changing their properties. If it is the case, galaxies in the transformational phase from star-forming to quiescent, such as mergers, SMGs, and AGNs, could spend this phase in intermediate matter densities. This can also explain the increase of QGs fractions with the density contrast, in the sense that galaxies quench their star formation by the time they arrive in higher density regions. 
A possible conflict of this scenario is the presence of AGNs in higher density regions at $\Delta_\text{local}\geq4$. 
We may interpret it in terms of different timescales for quenching star formation in galaxies. 
According to the merger-driven quenching scenario, the merging of two star-forming galaxies leads to a significant circumnuclear starburst. It simultaneously causes the gas inflow to the supermassive black holes and feeds an AGN in the center of the galaxy. The AGN feedback results in the rapid heating and removal of the cold gas and consequently quenching of star-formation within the galaxy (e.g., \cite{Sanders96,Springel05a,Springel05b,Hopkins06,Hopkins08,Yuan10,Alexander12}). 
Therefore, given this scenario, the AGN phase comes after the dusty star-forming phase, and it occurs just before reaching a quiescent phase. The different timescales after each phase of mergers, SMGs, and AGNs to transform into the QG phase (hereafter, referred to as quenching time) has been suggested in the literature. The quenching time inferred for SMGs --- after mergers and subsequent enhanced star formation ---  has been suggested as a few Gyr, whereas the post-AGN timescale has been inferred to be $\lesssim 1$ Gyr (e.g., \cite{Hopkins08,Schawinski09,Rodriguez19}). If this is the case with our samples, galaxies may traverse through intermediate matter density regions when they are undergoing the merger and SMGs phases but then migrate to higher density regions while changing their phase to AGNs. 

This second situation allows further insight into the process of quenching. The lack of galaxies in the early transforming phase of galaxy evolution at higher densities is reminiscent of ``pre-processed'' quenching of galaxies in clusters.
\citet{Zabludoff98} reported that a dominant fraction of elliptical galaxies in clusters may form via mergers in group environments before they accrete into clusters. Subsequent theoretical and observational studies have also claimed that physical mechanisms, such as mergers and ram pressure stripping, operating in intermediate density environments (e.g., groups, filaments, subclusters) are responsible for terminating star formation of galaxies in clusters (e.g., \cite{Fujita04,DeLucia12,Vijayaraghavan13,Taranu14,Pallero19}). This phenomenon is referred to as pre-processing. 
Recently, \citet{McNab21} have found no significant excess of transition galaxies -- post-starburst, blue quiescent, and green valley galaxies -- of the high-mass regime ($\Mstar>10^{10.5}$~M$_\odot$) in clusters at $z\sim 1$. To interpret these findings, they have suggested quenching via pre-processing for high-mass galaxies. 
Assuming this is also the case with our samples, galaxies may undergo merger-driven starbursts in regions with a local matter density contrast of $\Delta_\text{local}<4$, corresponding to fields and protocluster outskirts, and terminate their star formation before or during accretion onto high-density regions.
Consequently, mergers and SMGs would not possibly be seen in higher-density regions, whereas QGs and AGNs fractions increase with the matter density.

\section{Summary}
\label{sec:sum}

This study demonstrated the dependence of galactic properties, specifically star formation activities (i.e., \textit{SFR} and \textit{sSFR}) and galaxy types, on the environment and the stellar mass for galaxies at $z=2$. For the first time, we investigated the environment around galaxies by directly referring to the estimated matter density instead of using a smoothed galaxy overdensity field.
The matter density was evaluated from galaxy catalogs and Ly$\alpha$ forest observations by applying numerical reconstruction techniques. 
Our reconstructed density filed allowed us for the first time to study the relationship between galaxies and underdensities at $\Delta_\text{local}<1$ in detail. 
Using the reconstructed density field, COSTCO, we drew the following conclusions: 
\begin{enumerate}
\item We found the lack of dependence of star formation activities on the matter density contrast ($\Delta_\text{local}$) over uncertainties for both low-mass ($\Mstar<10^{10.5}$~M$_\odot$) and high-mass galaxies ($\Mstar\geq10^{10.5}$~M$_\odot$).
Nonetheless, it is still notable that we detect a consistent pattern of decreased star formation activities for high-mass galaxies in higher density contrast at $\Delta_\text{local}\geq4$ from all analyses, including rebinning, even though it is just from the median value. However, that statistical significance is low due to the large uncertainties. A larger sample with future surveys will provide solid conclusions with high significance.

\vspace{0.5cm}
\item We examined the fraction of mergers, SMGs, AGNs, and QGs relative to the density contrast.
Although our results in the above special galaxy types are still tentative with large uncertainties, we obtained the following trends. 
For mergers and SMGs, we see no clear dependence on the density contrast but find a lack of those types at higher density regions of $\Delta_\text{local}\geq4$. 
The probability of accidental occurrence of the lack of those samples at higher density regions was found to be low. 
Those results could indicate that mergers and SMGs tend to avoid higher-density regions. 
In contrast, AGN and QG fractions were found to increase with increasing the density contrast.
Considering all results from mergers, SMGs, AGNs, and QGs, we may infer a scenario in which galaxies evolve by transforming their properties while they migrate through the various density regimes of the cosmic web. In addition, the lack of mergers and SMGs may also imply the pre-processing of galaxy quenching.
\vspace{0.5cm}
\item We also studied the relation between galactic properties and the stellar mass ($\Mstar$).
The \textit{SFR} and \textit{sSFR} for a fixed $\Mstar$ bin were nearly the same across all $\Delta$ subsamples within their uncertainties.
A possible exception was detected for the high-mass regime at $\Mstar\geq10^{11}$~M$_\odot$ of the high-$\Delta_\text{local}$ subsample ($\Delta_\text{local}\geq4$), though its significance is still low.
In terms of mergers, SMGs, AGN, and QGs, we found a significant correlation between their fractions and $\Mstar$ for all galaxy types, unlike the density dependence. This could be another hint of mass-driven galaxy evolution.
\end{enumerate}

As we already argued, mass quenching must be the main driver of our galaxies at $z=2$. Nonetheless, environmental quenching cannot be completely ruled out. Although statistical significance is low due to a single data point, we still tentatively detected the sudden drop in star formation activities for high-mass galaxies in higher-density regions. We also tentatively found the change in the fraction of mergers, SMGs, AGNs, and QG with the local matter density. If those detections were real, they may indicate the environmental quenching described well by ``overconsumption'' and ``pre-processing'' models. 
Our galaxies might undergo mergers and subsequent starbursts at the intermediate density while traveling to the high density and gradually quench their star formation, known as pre-processing. Besides, high-mass galaxies in higher-density environments, $\Delta_\text{local}\geq4-6$ in case of our results, could have additional effects on the environment, called overconsumption, and accelerate their quenching.
More solid conclusions would be drawn from the complete galaxy samples in future surveys, such as by the Prime Focus Spectrograph (PFS; \cite{Sugai15}) on the Subaru telescope.
This next-generation facility should also enable more detailed studies involving the anisotropic cosmic web, including possible constraints on high-redshift galaxy intrinsic alignments to the cosmic web \citep{Krolewski17,Zhang22}.

\begin{ack}
We are grateful to  
Dr. Andrea Silva for kindly sharing the merger catalogs and 
Drs. Louise Edwards, Nima Chartab, Andrew B. Newman, and Mahdi Qezlou for helpful discussions.
We also thank the attendees of the 2023 KITP Cosmic Web Conference, \textit{The Co-evolution of the Cosmic Web and Galaxies across Cosmic Time}, including Drs. Chris Byrohl, Farhanul Hasan, Clotilde Laigle, Jounghun Lee, Hyunbae Park, and Hyunmi Song.
We also appreciate an anonymous referee for their valuable comments on improving our manuscript. 
R.M. acknowledges a Japan Society for the Promotion of Science (JSPS) Fellowship at Japan.  
This work is supported by the JSPS KAKENHI grant Nos. JP18J40088 and JP21H04490 (R.M.), JP18H058681418, JP19K14755 (K.G.L.), and JP21K13911 (M.A.). 
MA was supported by Simons Collaboration on ``Learning the Universe''.
Kavli IPMU was established by World Premier International Research Center Initiative (WPI), MEXT, Japan. 
Some of the data presented herein were obtained at the W.M. Keck Observatory, which is operated as a scientific partnership among the California Institute of Technology, the University of California and the National Aeronautics and Space Administration (NASA). The Observatory was made possible by the generous financial support of the W.M. Keck Foundation. The authors also wish to recognize and acknowledge the very significant cultural role and reverence that the summit of Maunakea has always had within the indigenous Hawai’ian community.
\end{ack}

\appendix 

\section{Tests for the Dependence on Stellar Mass Threshold}
\label{sec:app_dDM_Ms}

%
\begin{figure}[t]
\includegraphics[width=\linewidth]{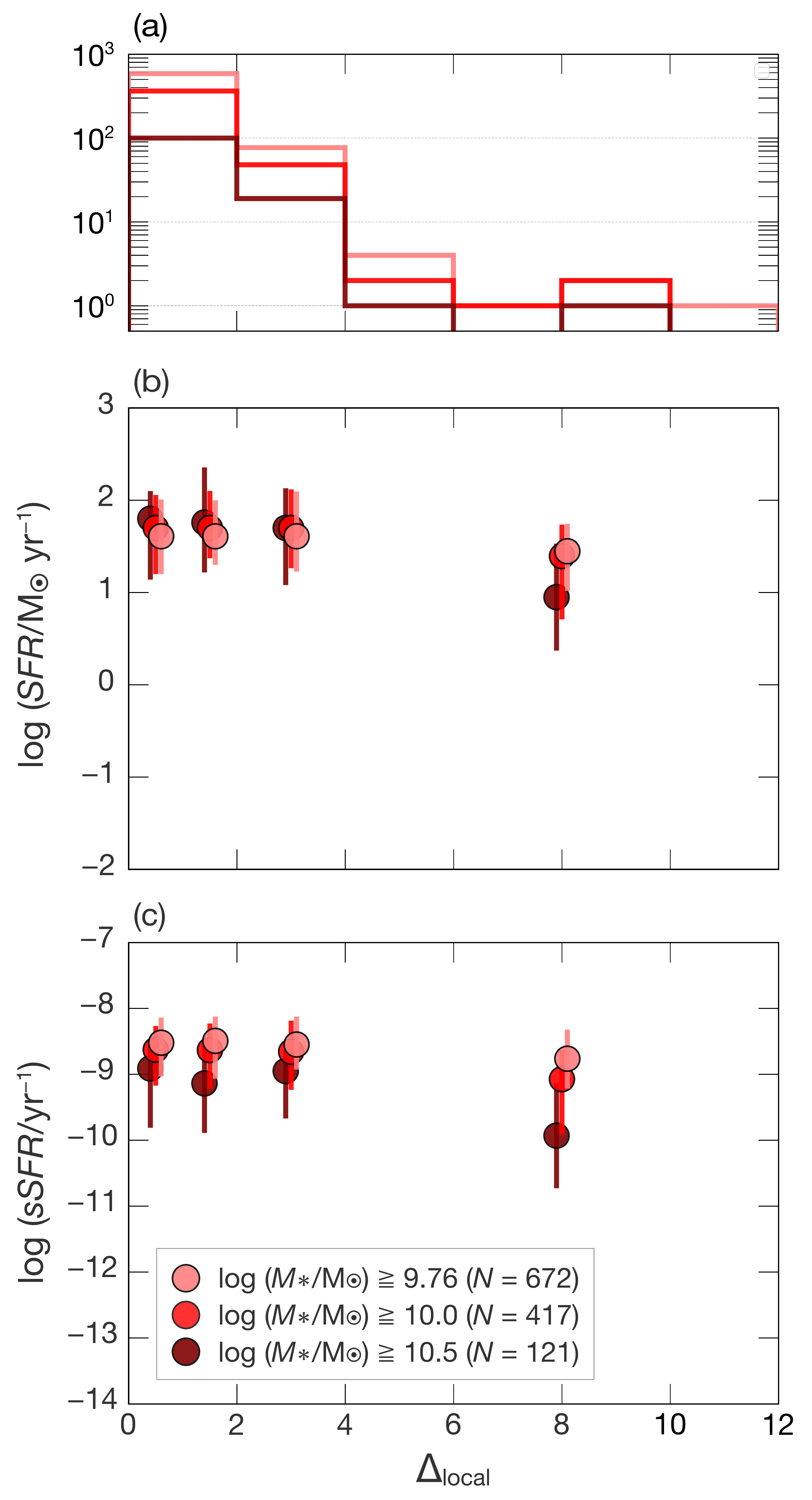}
\caption{
(a) Number histograms of stellar mass subsamples in each $\Delta_\text{local}=2$ bin.
The median (b) \textit{SFR} and (c) \textit{sSFR} of high-mass galaxies as a function of matter overdensity. The $\Delta_\text{local}$ bin size is [1, 1, 2, 8] to increase the sample size in each $\Mstar$ bin. The 16th and 84th percentiles are used as errorbars.
Different colors represent different stellar mass thresholds for mass subsamples.
}
\label{fig:SFA_dDM_MsComp}
\end{figure}

We test the impact on the stellar mass threshold for the results on high-mass galaxies by changing it as $\log{\Mstar/\text{M}_\odot}=10.50, 10.00, 9.76$. Note that the number of high-mass samples in these cases is ($121$, $417$, $672$) for $\log{\Mstar/\text{M}_\odot}=10.50, 10.00, 9.76$, in which the sample size for $\log{\Mstar/\text{M}_\odot}=9.76$ is half of the all of our sample. We use the same bin size as Figure~\ref{fig:SFA_dDM_rebin1} to verify the trend. The results on star formation activities on $\Delta_\text{local}$ for high-mass subsample are shown in Figure~\ref{fig:SFA_dDM_MsComp}. 
Both subsamples for $\log{\Mstar/\text{M}_\odot}=10.00$ and $9.76$ show consistency on \textit{SFR} and \textit{sSFR} with each other. In addition, there is no significant dependence on $\Delta_\text{local}$ beyond the uncertainties. Nonetheless, their median values show a slight decline, as we saw in the case of high-mass galaxies with the threshold of $\log{\Mstar/\text{M}_\odot}=10.50$. Their star formation activities decrease by $0.30-0.40$ and $0.16-0.22$ dex, for $\log{\Mstar/\text{M}_\odot}=10.00$ and $9.76$ subsamples, respectively.

These tests reveal a qualitatively similar trend for high-mass galaxies in the relationship between star formation activities and $\Delta_\text{local}$, regardless of the $\Mstar$ thresholds, at least those utilized in the test. We find no significant dependence of star formation activities on $\Delta_\text{local}$ beyond uncertainties. Nonetheless, it would be interesting that all subsamples detect a decline of their median values at $\Delta_\text{local}\geq4.0$. Given the extent of decrease for $10.00$ and $9.76$ thresholds being smaller compared to the subsample with a $10.50$ threshold, using a $10.50$ threshold could help to confirm a trend clearly if that presents in the star formation activities and $\Delta_\text{local}$ relations.

\section{Test for the Variation on $\Delta_\text{local}$ Criteria}
\label{sec:app_delta}

%
\begin{figure}[t]
\includegraphics[width=\linewidth]{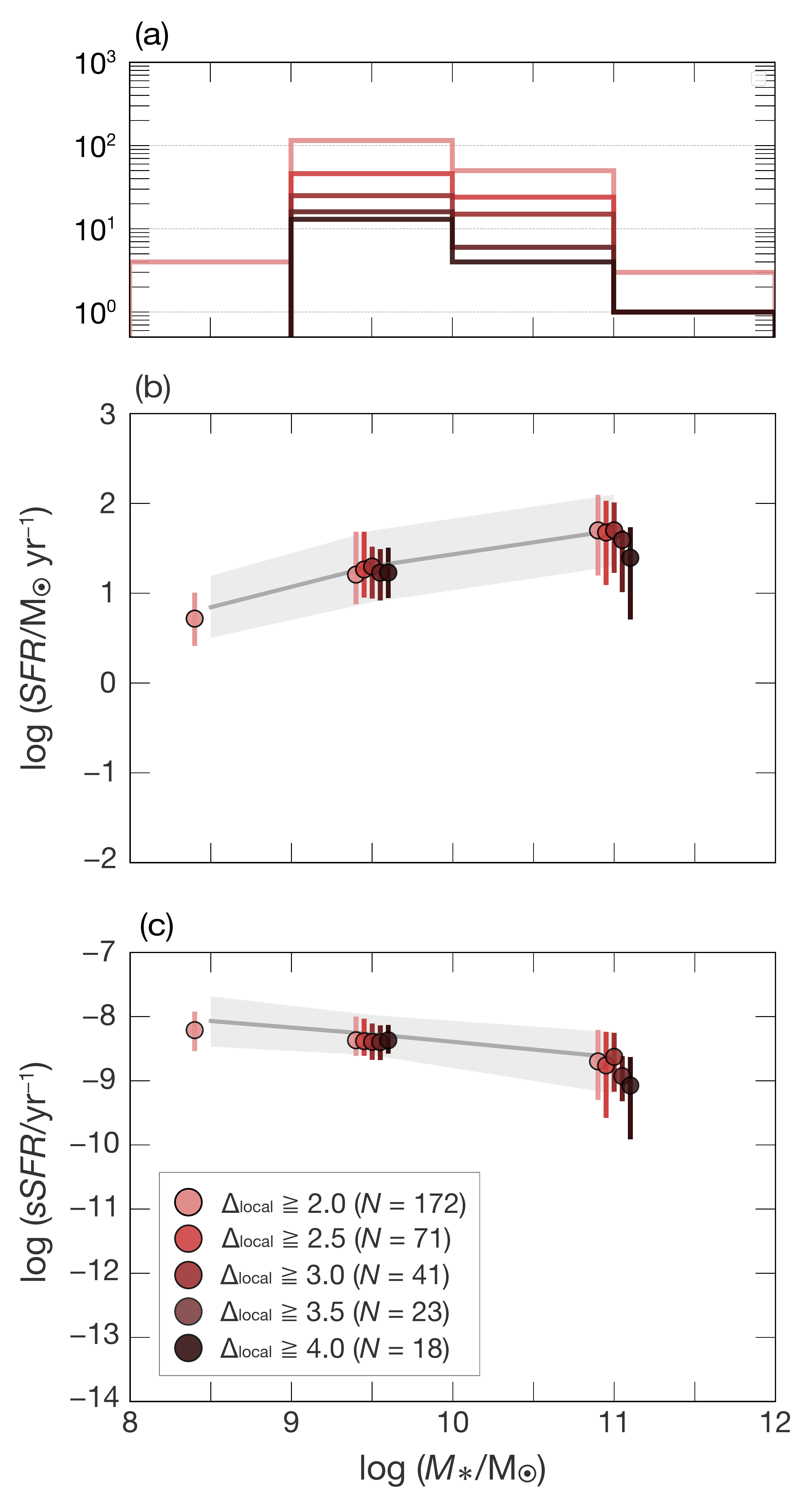}
\caption{
(a) Number histograms of matter density subsamples in each $\Mstar=10^1$~M$_\odot$ bin from $\Mstar=10^8$ to $10^{12}$~M$_\odot$. 
The median (b) \textit{SFR} and (c) \textit{sSFR} to $\Mstar$ in $\log{\Mstar}$ bin sizes of [$1,1,2$]. The error bars indicate the percentiles 16th and 84th of \textit{SFR} and \textit{sSFR}. A gray line and shades are the medians and 68 percentiles of \textit{SFR} and \textit{sSFR} of the entire mass-completeness sample. Data points are slightly shifted horizontally for clarity. Different colors represent different $\Delta$ thresholds for matter density subsamples.
}
\label{fig:SFA_Ms_dDMComp}
\end{figure}

In Figure~\ref{fig:SFA_Ms_dDMComp}, we test the impact on the matter density threshold for the results on high-$\Delta_\text{local}$ subsample. We used the threshold for the high-$\Delta_\text{local}$ subsample of $\Delta_\text{local}\geq2.0$, $2.5$, $3.0$, $3.5$, $4.0$ whose sample sizes are ($172$, $71$, $41$, $23$, $18$).
All subsamples are consistent with the entire median, as shown by the grey line within the uncertainties. The possible decline of medians is confirmed only in $\Delta_\text{local}\geq3.5$ and $4.0$ subsamples, which are $0.1-0.3$ dex and $0.3-0.4$ dex in \textit{SFR} and \textit{sSFR} to the entire median.

These tests show that the matter density threshold could influence the results on the high-$\Delta_\text{local}$ subsample. The decline of star formation activities in high-mass galaxies in the highest density region, as commonly seen in the main text, though its statistical significance is low, is only observed when we apply the threshold $\Delta_\text{local}\geq3.5$. If our results reflect a true relation of galaxy properties on the matter density contrast in the universe, the regions with $\Delta_\text{local}\geq3.5$ might be special at $z\sim2$ as causing the environmental quenching. Hence, it might be intriguing to study such regions in more detail in the context of galaxy evolution together with the cosmic web utilizing upcoming galaxy survey data.

\bibliographystyle{aasjournal}

\begin{thebibliography}{120}
\expandafter\ifx\csname natexlab\endcsname\relax\def\natexlab#1{#1}\fi

\bibitem[{
{Alexander} \& {Hickox}(2012)
{Alexander}, D.~M.,\& {Hickox}, R.~C.
}]{Alexander12}
{Alexander D.~M.} \& {Hickox R.~C.}, 2012, \nat, 56, 93

\bibitem[{
{{\'A}lvarez Crespo}{ et al.}(2021)
{{\'A}lvarez Crespo}, N.,{Smoli{\'c}}, V.,{Finoguenov}, A.,{Barrufet}, L.,\&{Aravena}, M.
}]{Crespo21}
{{\'A}lvarez Crespo}, N., {et~al.} 2021, \aap, 646, 174

\bibitem[{
{Ando}{ et al.}(2020)
{Ando},{Makoto},{Shimasaku},{Kazuhiro},\&{Momose},{Rieko}
}]{Ando20}
{Ando M.}, {Shimasaku K.} \& {Momose R.} 2020, \mnras, 496, 3169

\bibitem[{
{Ata}{ et al.}(2022)
{Ata}, Metin, {Lee}, Khee-Gan, {Vecchia}, Claudio Dalla, {Kitaura}, Francisco-Shu,{Cucciati}, Olga, {Lemaux}, Brian C.,{Kashino}, Daichi, \&{M{\"u}ller}, Thomas
}]{Ata22}
{Ata M.}, {et~al.} 2022, \nat, 6, 857

\bibitem[{
{Ata}{ et al.}(2021)
{Ata}, Metin, {Kitaura}, Francisco-Shu, {Lee}, Khee-Gan, {Lemaux}, Brian C., {Kashino}, Daichi, {Cucciati}, Olga, {Hern{\'a}ndez-S{\'a}nchez}, M{\'o}nica, \&{Le F{\`e}vre}, Oliver
}]{Ata21}
{Ata M.}, {et~al.} 2021, \mnras, 500, 3194

\bibitem[{
{Baldry}{ et al.}(2004)
{Baldry}, Ivan K., {Glazebrook}, Karl, {Brinkmann}, Jon, {Ivezi{\'c}}, {\v{Z}}eljko, {Lupton}, Robert H., {Nichol}, Robert C.,\& {Szalay}, Alexander S.
}]{Baldry04}
{Baldry I.~K.}, {et~al.} 2004, \apj, 600, 681

\bibitem[{
{Balogh}{ et al.}(2016)
{Balogh}, Michael L.,{McGee}, Sean L.,{Mok}, Angus,{Muzzin}, Adam,{van der Burg}, Remco F.~J.,{Bower}, Richard G.,{Finoguenov}, Alexis,{Hoekstra}, Henk,{Lidman}, Chris,{Mulchaey}, John S.,{Noble}, Allison,{Parker}, Laura C.,{Tanaka}, Masayuki,{Wilman}, David J.,{Webb}, Tracy,{Wilson}, Gillian,\&{Yee}, Howard K.~C.
}]{Balogh16}
{Balogh M.~L.}, {et~al.} 2016, \mnras, 456, 4364

\bibitem[{
{Balogh}{ et al.}(2000)
{Balogh}, Michael L., {Navarro}, Julio F.,\& {Morris}, Simon L.
}]{Balogh00}
{Balogh M.~L.}, {Navarro J.~F.} \& {Morris S.~L.} 2000, \apj, 540, 113

\bibitem[{
{Bekki}(2014)
{Bekki}, Kenji
}]{Bekki14}
{Bekki K.}, 2014, \mnras, 438, 444

\bibitem[
{Behroozi}{ et al.}(2019)
{Behroozi}, Peter, {Wechsler}, Risa H., {Hearin}, Andrew P., \& {Conroy}, Charlie
]{Behroozi19}
{Behroozi P.}, {et~al.}, 2019, \mnras, 488, 3143

\bibitem[
{Behroozi}{ et al.}(2013)
{Behroozi}, Peter S., {Wechsler}, Risa H., \& {Conroy}, Charlie
]{Behroozi13}
{Behroozi P.~S.}, {Wechsler R.~H.}, \& {Conroy C.}, 2013, \apj, 770, 57

\bibitem[{
{Blanton} \& {Moustakas}(2009)
{Blanton}, Michael R.,\& {Moustakas}, John
}]{Blanton09}
{Blanton M.~R.} \& {Moustakas J.} 2009, \araa, 47, 159

\bibitem[{
{Blanton}{ et al.}(2005)
{Blanton}, Michael R., {Eisenstein}, Daniel, {Hogg}, David W., {Schlegel}, David J.,\& {Brinkmann}, J.
}]{Blanton05}
{Blanton M.~R.}, {et~al.} 2005, \apj, 629, 143

\bibitem[{
{Blanton}{ et al.}(2003)
{Blanton}, Michael R., {Hogg}, David W., {Bahcall}, Neta A., {Baldry}, Ivan K., {Brinkmann}, J., {Csabai}, Istv{\'a}n, {Eisenstein}, Daniel, {Fukugita}, Masataka , {Gunn}, James E., {Ivezi{\'c}}, {\v{Z}}eljko, {Lamb}, D.~Q., {Lupton}, Robert H., {Loveday}, Jon, {Munn}, Jeffrey A., {Nichol}, R.~C., {Okamura}, Sadanori, {Schlegel}, David J., {Shimasaku}, Kazuhiro, {Strauss}, Michael A., {Vogeley}, Michael S.,\& {Weinberg}, David H.
}]{Blanton03}
{Blanton M.~R.}, {et~al.} 2003, \apj, 594, 186

\bibitem[{
{Boselli}{ et al.}(2022)
{Boselli}, Alessandro,{Fossati}, Matteo,\&{Sun}, Ming
}]{Boselli22}
{Boselli A.}, {Fossati M.} \& {Sun M.} 2022, \aapr, 30, 3

\bibitem[{
{Brammer}{ et al.}(2009)
{Brammer}, G.~B., {Whitaker}, K.~E., {van Dokkum}, P.~G., {Marchesini}, D., {Labb{\'e}}, I., {Franx}, M., {Kriek}, M., {Quadri}, R.~F., {Illingworth}, G., {Lee}, K. -S., {Muzzin}, A.,\& {Rudnick}, G.
}]{Brammer09}
{Brammer G.~B.}, {et~al.} 2009, \apjl, 706, L173

\bibitem[{
{Brisbin}{ et al.}(2017)
{Brisbin}, Drew, {Miettinen}, Oskari,{Aravena}, Manuel, {Smol{\v{c}}i{\'c}}, Vernesa, {Delvecchio}, Ivan, {Jiang}, Chunyan, {Magnelli}, Benjamin, {Albrecht}, Marcus,{Arancibia}, Alejandra Mu{\~n}oz, {Aussel}, Herv{\'e}, {Baran}, Nikola, {Bertoldi}, Frank, {B{\'e}thermin}, Matthieu, {Capak}, Peter, {Casey}, Caitlin M., {Civano}, Francesca, {Hayward}, Christopher C.,{Ilbert}, Olivier, {Karim}, Alexander, {Le Fevre}, Olivier, {Marchesi}, Stefano, {McCracken}, Henry Joy, {Navarrete}, Felipe, {Novak}, Mladen, {Riechers}, Dominik, {Padilla}, Nelson, {Salvato}, Mara, {Scott}, Kimberly, {Schinnerer}, Eva, {Sheth}, Kartik, \&{Tasca}, Lidia
}]{Brisbin17}
{Brisbin D.}, {et~al.} 2017, \aap, 608, 15

\bibitem[{
{Brusa}{ et al.}(2009)
{Brusa}, M.,{Fiore}, F.,{Santini}, P.,{Grazian}, A.,{Comastri}, A.,{Zamorani}, G.,{Hasinger}, G.,{Merloni}, A. ,{Civano}, F.,{Fontana}, A.,\&{Mainieri}, V.
}]{Brusa09}
{Brusa M.}, {et~al.} 2007, \aap, 507, 1277

\bibitem[{
{Bundy}{ et al.}(2009)
{Bundy},{Kevin},{Fukugita},{Masataka},{Ellis},{Richard S.},{Targett},{Thomas A.},{Belli},{Sirio},\&{Kodama},{Tadayuki}
}]{Bundy09}
{Bundy K.}, {et~al.} 2009, \apj, 697, 1369

\bibitem[{
{Capak}{ et al.}(2007)
{Capak},P.,{Aussel},H.,{Ajiki},M.,{McCracken},H.~J.,{Mobasher},B.,{Scoville},N.,{Shopbell},P.,{Taniguchi},Y.,{Thompson},D.,{Tribiano},S.,{Sasaki},S.,{Blain},A.~W.,{Brusa},M.,{Carilli},C.,{Comastri},A.,{Carollo},C.~M.,{Cassata},P.{Colbert},J.,{Ellis},R.~S.,{Elvis},M.,{Giavalisco},M.,{Green},W.,{Guzzo},L.,{Hasinger},G.,{Ilbert},O.,{Impey},C.,{Jahnke},K.,{Kartaltepe},J.,{Kneib},J. -P.,{Koda},J.,{Koekemoer},A.,{Komiyama},Y.,{Leauthaud},A.,{Le Fevre},O.,{Lilly},S.,{Liu},C.,{Massey},R.,{Miyazaki},S.,{Murayama},T.,{Nagao},T.,{Peacock},J.~A.,{Pickles},A.,{Porciani},C.,{Renzini},A.,{Rhodes},J.,{Rich},M.,{Salvato},M.,{Sanders},D.~B.,{Scarlata},C.,{Schiminovich},D.,{Schinnerer},E.,{Scodeggio},M.,{Sheth},K.,{Shioya},Y.,{Tasca},L.~A.~M.,{Taylor},J.~E.,{Yan},L.,\&{Zamorani},G.
}]{Capak07}
{Capak E.}, {et~al.} 2007, \apjs, 172, 99

\bibitem[{
{Casey}(2016)
{Casey},{Caitlin M.}
}]{Casey16}
{Casey C.~M.}, 2016, \apj, 824, 36

\bibitem[{
{Champagne}{ et al.}(2021)
{Champagne}, Jaclyn B., {Casey}, Caitlin M., {Zavala}, Jorge A., {Cooray}, Asantha, {Dannerbauer}, Helmut, {Fabian}, Andrew, {Hayward}, Christopher C., {Long}, Arianna S., \&{Spilker}, Justin S.
}]{Champagne21}
{Champagne J.~B.}, {et~al.} 2021, \apj, 913, 110

\bibitem[
{Chang}{ et al.}(2022)
{Chang}, W., {Fang}, G.,{Gu}, Y.,{Lin}, Z., {Lu}, S., \&{Kong}, X.
]{Chang22}
{Chang W.}, {et~al.} 2022, \apj, 936, 47

\bibitem[{
{Chapman}{ et al.}(2009)
{Chapman}, S.~C.,{Blain}, A.,{Ibata}, R.,{Ivison}, R.~J.,{Smail}, I.,\&{Morrison}, G.
}]{Chapman09}
{Chapman S.~C.}, {et~al.} 2009, \apj, 691, 560

\bibitem[{
{Chartab}{ et al.}(2020)
{Chartab}, Nima,{Mobasher}, Bahram,{Darvish}, Behnam,{Finkelstein}, Steve,{Guo}, Yicheng,{Kodra}, Dritan,{Lee}, Kyoung-Soo,{Newman}, Jeffrey A.,{Pacifici}, Camilla,{Papovich}, Casey,{Sattari}, Zahra,{Shahidi}, Abtin,{Dickinson}, Mark E.,{Faber}, Sandra M.,{Ferguson}, Henry C.,{Giavalisco}, Mauro,\&{Jafariyazani}, Marziye
}]{Chartab20}
{Chartab N.}, {et~al.} 2020, \apj, 890, 7

\bibitem[{
{Chuter}{ et al.}(2011)
{Chuter}, R.~W.,{Almaini}, O.,{Hartley}, W.~G.,{McLure}, R.~J.,{Dunlop}, J.~S.,{Foucaud}, S.,{Conselice}, C.~J.,{Simpson}, C.,{Cirasuolo}, M.,\&{Bradshaw}, E.~J.
}]{Chuter11}
{Chuter R.~W.}, {et~al.} 2011, \mnras, 413, 1678

\bibitem[{
{Cowley}{ et al.}(2016)
{Cowley}, Michael J., {Spitler}, Lee R., {Tran}, Kim-Vy H., {Rees}, Glen A., {Labb{\'e}}, Ivo, {Allen}, Rebecca J., {Brammer}, Gabriel B., {Glazebrook}, Karl, {Hopkins}, Andrew M., {Juneau}, St{\'e}phanie, {Kacprzak}, Glenn G., {Mullaney}, James R., {Nanayakkara}, Themiya, {Papovich}, Casey, {Quadri}, Ryan F., {Straatman}, Caroline M.~S., {Tomczak}, Adam R., \&{van Dokkum}, Pieter G.
}]{Cowley16}
{Cowley M.~J.}, {et~al.} 2016, \mnras, 457, 629

\bibitem[{
{Croton}{ et al.}(2006)
{Croton}, Darren J., {Springel}, Volker, {White}, Simon D.~M., {De Lucia}, G., {Frenk}, C.~S., {Gao}, L., {Jenkins}, A., {Kauffmann}, G., {Navarro}, J.~F.,\& {Yoshida}, N.
}]{Croton06}
{Croton D.~J.}, {et~al.} 2006, \mnras, 365, 11
 
\bibitem[{
{Daddi}{ et al.}(2007)
{Daddi},E.,{Dickinson},M.,{Morrison},G.,{Chary},R.,{Cimatti},A.,{Elbaz},D.,{Frayer},D.,{Renzini},A.,{Pope},A.,{Alexander},D.~M.,{Bauer},F.~E.,{Giavalisco},M.,{Huynh},M.,{Kurk},J.,\&{Mignoli},M.
}]{Daddi07}
{Daddi E.}, {et~al.} 2007, \apj, 670, 156

\bibitem[{
{Dalla Vecchia} \& {Schaye}(2008)
{Dalla Vecchia}, Claudio,\& {Schaye}, Joop
}]{DallaVecchia08}
{Dalla Vecchia C.} \& {Schaye J.} 2008, \mnras, 387, 1431

\bibitem[{
{Darvish}{ et al.}(2016)
{Darvish}, Behnam, {Mobasher}, Bahram, {Sobral}, David, {Rettura}, Alessandro, {Scoville}, Nick, {Faisst}, Andreas,\&{Capak}, Peter
}]{Darvish16}
{Darvish B.}, {et~al.} 2016, \apj, 825, 113

\bibitem[{
{De Lucia}{ et al.}(2012)
{De Lucia}, Gabriella, {Weinmann}, Simone, {Poggianti}, Bianca M., {Arag{\'o}n-Salamanca}, Alfonso, \&{Zaritsky}, Dennis
}]{DeLucia12}
{De Lucia G.}, {et~al.} 2012, \mnras, 423, 1277

\bibitem[{
{Dekel} \& {Silk}(1986)
{Dekel}, A.,\& {Silk}, J.
}]{Dekel86}
{Dekel A.} \& {Silk J.} 1986, \apj, 303, 39

\bibitem[{
{Delahaye}{ et al.}(2017)
{Delahaye}, A.~G.,{Webb}, T.~M.~A.,{Nantais}, J.,{DeGroot}, A.,{Wilson}, G.,{Muzzin}, A.,{Yee}, H.~K.~C.,{Foltz}, R.,{Noble}, A.~G.,{Demarco}, R.,{Tudorica}, A.,{Cooper}, M.~C.,{Lidman}, C.,{Perlmutter}, S.,{Hayden}, B.,{Boone}, K.,\&{Surace}, J.
}]{Delahaye17}
{Delahaye A.~G.}, {et~al.} 2017, \apj, 843, 126

\bibitem[{
{Delvecchio}{ et al.}(2017)
{Delvecchio},I.,{Smol{\v{c}}i{\'c}},V.,{Zamorani},G.,{Lagos},C. Del P.,{Berta},S.,{Delhaize},J.,{Baran},N.,{Alexander},D.~M.,{Rosario},D.~J.,{Gonzalez-Perez},V.,{Ilbert},O.,{Lacey},C.~G.,{Le F{\`e}vre},O.,{Miettinen},O.,{Aravena},M.,{Bondi},M.,{Carilli},C.,{Ciliegi},P.,{Mooley},K.,{Novak},M.,{Schinnerer},E.,{Capak},P.,{Civano},F.,{Fanidakis},N.,{Herrera Ruiz},N.,{Karim},A.,{Laigle},C.,{Marchesi},S.,{McCracken},H.~J.,{Middleberg},E.,{Salvato},M.,\&{Tasca},L.
}]{Delvecchio17}
{Delvecchio I.}, {et~al.} 2017, \aap, 602, 3

\bibitem[{
{Digby-North}{ et al.}(2010)
{Digby-North}, J.~A.,{Nandra}, K.,{Laird}, E.~S.,{Steidel}, C.~C.,{Georgakakis}, A.,{Bogosavljevi{\'c}}, M.,{Erb}, D.~K.,{Shapley}, A.~E.,{Reddy}, N.~A.,\&{Aird}, J.
}]{Digby-North10}
{Digby-North J.~A.}, {et~al.} 2010, \mnras, 407, 846

\bibitem[{
{Dong}{ et al.}(2023)
{Dong},{Chenze},{Lee},{Khee-Gan},{Ata},{Metin},{Horowitz},{Benjamin},\&{Momose},{Rieko}
}]{Dong23}
{Dong C.}, {et~al.} 2023, \apjl, 945, L28

\bibitem[{
{Fabian}(2012)
{Fabian}, A.~C.
}]{Fabian12}
{Fabian A.~C.}, 2012, \araa, 50, 455

\bibitem[{
{Fang}{ et al.}(2013)
{Fang}, Jerome J., {Faber}, S.~M., {Koo}, David C., \&{Dekel}, Avishai
}]{Fang13}
{Fang J.~J.}, {et~al.} 2013, \apj, 776, 63

\bibitem[{
{Fossati}{ et al.}(2017)
{Fossati}, M.,{Wilman}, D.~J.,{Mendel}, J.~T.,{Saglia}, R.~P.,{Galametz}, A.,{Beifiori}, A.,{Bender}, R.,{Chan}, J.~C.~C.,{Fabricius}, M.,{Bandara}, K.,{Brammer}, G.~B.,{Davies}, R.,{F{\"o}rster Schreiber}, N.~M.,{Genzel}, R.,{Hartley}, W.,{Kulkarni}, S.~K.,{Lang}, P.,{Momcheva}, I.~G.,{Nelson}, E.~J.,{Skelton}, R.,{Tacconi}, L.~J.,{Tadaki}, K.,{{\"U}bler}, H. ,{van Dokkum}, P.~G.,{Wisnioski}, E.,{Whitaker}, K.~E.,{Wuyts}, E.,\&{Wuyts}, S.
}]{Fossati17}
{Fossati M.}, {et~al.} 2017, \apj, 835, 153

\bibitem[{
{Fujita}(2004)
{Fujita},Yutaka
}]{Fujita04}
{Fujita Y.}, 2004, \pasj, 56, 29

\bibitem[{
{Gonz{\'a}lez}{ et al.}(2011)
{Gonz{\'a}lez},{Juan E.},{Lacey},{C.~G.},{Baugh},{C.~M.},\&{Frenk},{C.~S.}
}]{Gonzalez11}
{Gonz{\'a}lez J.~E.}, {et~al.} 2011, \mnras, 413, 749

\bibitem[{
{Gunn} \& {Gott}(1972)
{Gunn}, James E., \&{Gott}, J. Richard, III
}]{Gunn72}
{Gunn J.~E.} \& {Gott J.~R. III} 1972, \apj, 176, 1

\bibitem[{
{Guo}{ et al.}(2017)
{Guo}, Yicheng,{Bell}, Eric F.,{Lu}, Yu,{Koo}, David C.,{Faber}, S.~M.,{Koekemoer}, Anton M.,{Kurczynski}, Peter,{Lee}, Seong-Kook,{Papovich}, Casey,{Chen}, Zhu,{Dekel}, Avishai,{Ferguson}, Henry C.,{Fontana}, Adriano,{Giavalisco}, Mauro,{Kocevski}, Dale D.,{Nayyeri}, Hooshang,{P{\'e}rez-Gonz{\'a}lez}, Pablo G.,{Pforr}, Janine,{Rodr{\'\i}guez-Puebla}, Aldo,\&{Santini}, Paola
}]{Guo17}
{Guo Y.}, {et~al.} 2017, \apjl, 841, L22

\bibitem[{
{Hashimoto}{ et al.}(2013)
{Hashimoto}, Takuya, {Ouchi}, Masami,{Shimasaku}, Kazuhiro, {Ono}, Yoshiaki,{Nakajima}, Kimihiko, {Rauch}, Michael, {Lee}, Janice, \&{Okamura}, Sadanori
}]{Hashimoto13}
{Hashimoto T.}, {et~al.} 2013, \apj, 765, 70

\bibitem[{
{Hickox}{ et al.}(2012)
{Hickox}, Ryan C.,{Wardlow}, J.~L.,{Smail}, Ian,{Myers}, A.~D.,{Alexander}, D.~M.,{Swinbank}, A.~M.,{Danielson}, A.~L.~R.,{Stott}, J.~P.,{Chapman}, S.~C.,{Coppin}, K.~E.~K.,{Dunlop}, J.~S.,{Gawiser}, E.,{Lutz}, D.,{van der Werf}, P.,\&{Wei{\ss}}, A.
}]{Hickox12}
{Hickox R.~C.}, {et~al.} 2012, \mnras, 421, 284

\bibitem[{
{Hine}{ et al.}(2016)
{Hine}, N.~K.,{Geach}, J.~E.,{Alexander}, D.~M.,{Lehmer}, B.~D.,{Chapman}, S.~C.,\&{Matsuda}, Y.
}]{Hine16}
{Hine N.~K.}, {et~al.} 2016, \mnras, 455, 2363

\bibitem[{
{Horowitz}{ et al.}(2019)
{Horowitz}, Benjamin, {Lee}, Khee-Gan, {White}, Martin, {Krolewski}, Alex, \&{Ata}, Metin
}]{Horowitz19}
{Horowitz B.}, {et~al.} 2019, \apj, 887, 61

\bibitem[{
{Horowitz}{ et al.}(2021)
{Horowitz}, Benjamin, {Zhang}, Benjamin, {Lee}, Khee-Gan, \&{Kooistra}, Robin
}]{Horowitz21}
{Horowitz B.}, {et~al.} 2021, \apj, 906, 110

\bibitem[{
{Hopkins}{ et al.}(2008)
{Hopkins}, Philip F., {Hernquist}, Lars, {Cox}, Thomas J., \&{Kere{\v{s}}}, Du{\v{s}}an
}]{Hopkins08}
{Hopkins P.~F.}, {et al.} 2008, \apjs, 175, 356

\bibitem[{
{Hopkins}{ et al.}(2006)
{Hopkins}, Philip F., {Somerville}, Rachel S., {Hernquist}, Lars, {Cox}, Thomas J.,{Robertson}, Brant, \&{Li}, Yuexing
}]{Hopkins06}
{Hopkins P.~F.}, {et~al.} 2006, \apj, 652, 864

\bibitem[{
{Huang}{ et al.}(2022)
{Huang},{Yun},{Lee},{Kyoung-Soo},{Cucciati},{Olga},{Lemaux},{Brian},{Sawicki},{Marcin},{Malavasi},{Nicola},{Ramakrishnan},{Vandana},{Xue},{Rui},{Cassara},{Letizia P.},{Chiang},{Yi-Kuan},{Dey},{Arjun},{Gwyn},{Stephen D.~J.},{Hathi},{Nimish},{Pentericci},{Laura},{Prescott},{Moire},\&{Zamorani},{Gianni}
}]{Huang22}
{Huang Y.}, {et~al.} 2022, \apj, 941, 134

\bibitem[{
{Ilbert}{ et al.}(2013)
{Ilbert}, O., {McCracken}, H.~J., {Le F{\`e}vre}, O., {Capak}, P., {Dunlop}, J.,{Karim}, A., {Renzini}, M.~A., {Caputi}, K., {Boissier}, S., {Arnouts}, S., {Aussel}, H., {Comparat}, J., {Guo}, Q., {Hudelot}, P., {Kartaltepe}, J., {Kneib}, J.~P., {Krogager}, J.~K., {Le Floc'h}, E., {Lilly}, S., {Mellier}, Y., {Milvang-Jensen}, B., {Moutard}, T., {Onodera}, M., {Richard}, J., {Salvato}, M., {Sanders}, D.~B., {Scoville}, N., {Silverman}, J.~D., {Taniguchi}, Y., {Tasca}, L., {Thomas}, R., {Toft}, S., {Tresse}, L., {Vergani}, D., {Wolk}, M., \&{Zirm}, A.
}]{Ilbert13}
{Ilbert O.}, {et~al.} 2013, \aap, 556, 55

\bibitem[{
{Ilbert}{ et al.}(2006)
{Ilbert},O.,{Arnouts},S.,{McCracken},H.~J.,{Bolzonella},M.,{Bertin},E.,{Le F{\`e}vre},O.,{Mellier},Y.,{Zamorani},G.,{Pell{\`o}},R.,{Iovino},A.,{Tresse},L.,{Le Brun},V.,{Bottini},D.,{Garilli},B.,{Maccagni},D.,{Picat},J.~P.,{Scaramella},R.,{Scodeggio},M.,{Vettolani},G.,{Zanichelli},A.,{Adami},C.,{Bardelli},S.,{Cappi},A.,{Charlot},S.,{Ciliegi},P.,{Contini},T.,{Cucciati},O.,{Foucaud},S.,{Franzetti},P.,{Gavignaud},I.,{Guzzo},L.,{Marano},B.,{Marinoni},C.,{Mazure},A.,{Meneux},B.,{Merighi},R.,{Paltani},S.,{Pollo},A.,{Pozzetti},L.,{Radovich},M.,{Zucca},E.,{Bondi},M.,{Bongiorno},A.,{Busarello},G.,{de La Torre},S.,{Gregorini},L.,{Lamareille},F.,{Mathez},G.,{Merluzzi},P.,{Ripepi},V.,{Rizzo},D.,\&{Vergani},D.
}]{Ilbert06}
{Ilbert O.}, {et~al.} 2006, \aap, 457, 841

\bibitem[{
{Ito}{ et al.}(2022)
{Ito},{Kei},{Tanaka},{Masayuki},{Miyaji},{Takamitsu},{Ilbert},{Olivier},{Kauffmann},{Olivier B.},{Koekemoer},{Anton M.},{Marchesi},{Stefano},{Shuntov},{Marko},{Toft},{Sune},{Valentino},{Francesco},\&{Weaver},{John R.}
}]{Ito22}
{Ito K.}, {et~al.} 2022, \apj, 929, 53

\bibitem[{
{Ito}{ et al.}(2021)
{Ito},{Kei},{Kashikawa},{Nobunari},{Tanaka},{Masayuki},{Kubo},{Mariko},{Liang},{Yongming},{Toshikawa},{Jun},{Uchiyama},{Hisakazu},{Ishimoto},{Rikako},{Yoshioka},{Takehiro},\&{Takeda},{Yoshihiro}
}]{Ito21}
{Ito K.}, {et~al.} 2021, \apj, 916, 35

\bibitem[{
{Jasche} \& {Wandelt}(2013)
{Jasche},{Jens},\&{Wandelt},{Benjamin D.}
}]{Jasche13}
{Jasche J.} \& {Wandelt B.~D.} 2013, \mnras, 432, 894

\bibitem[{
{Ji}{ et al.}(2018)
{Ji}, Zhiyuan,{Giavalisco}, Mauro,{Williams}, Christina C.,{Faber}, Sandra M.,{Ferguson}, Henry C.,{Guo}, Yicheng,{Liu}, Teng,\&{Lee}, Bomee
}]{Ji18}
{Ji Z.}, {et~al.} 2018, \apj, 862, 135
 
\bibitem[{
{Jones}{ et al.}(2017)
{Jones},{Suzy F.},{Blain},{Andrew W.},{Assef},{Roberto J.},{Eisenhardt},{Peter},{Lonsdale},{Carol},{Condon},{James},{Farrah},{Duncan},{Tsai},{Chao-Wei},{Bridge},{Carrie},{Wu},{Jingwen},{Wright},{Edward L.},\&{Jarrett},{Tom}
}]{Jones17}
{Jones S.~F.}, {et~al.} 2017, \mnras, 469, 4565

\bibitem[{
{Kashino}{ et al.}(2013)
{Kashino},D.,{Silverman},J.~D.,{Rodighiero},G.,{Renzini},A.,{Arimoto},N.,{Daddi},E.,{Lilly},S.~J.,{Sanders},D.~B.,{Kartaltepe},J. ,{Zahid},H.~J.,{Nagao},T.,{Sugiyama},N.,{Capak},P.,{Carollo},C.~M.,{Chu},J.,{Hasinger},G.,{Ilbert},O.,{Kajisawa},M.,{Kewley},L.~J.,{Koekemoer},A.~M.,{Kova{\v{c}}},K.,{Le F{\`e}vre},O.,{Masters},D.,{McCracken},H.~J.,{Onodera},M.,{Scoville},N.,{Strazzullo},V.,{Symeonidis},M.,\&{Taniguchi},Y.
}]{Kashino13}
{Kashino D.}, {et~al.} 2013, \apjl, 777, L8

\bibitem[{
{Kauffmann}{ et al.}(2004)
{Kauffmann},{Guinevere},{White},{Simon D.~M.},{Heckman},{Timothy M.},{M{\'e}nard},{Brice},{Brinchmann},{Jarle},{Charlot},{St{\'e}phane},{Tremonti},{Christy},\&{Brinkmann},{Jon}
}]{Kauffmann04}
{Kauffmann G.}, {et~al.} 2004, \mnras, 353, 713

\bibitem[{
{Kauffmann}{ et al.}(2003)
{Kauffmann}, Guinevere, {Heckman}, Timothy M., {White}, Simon D.~M., {Charlot}, St{\'e}phane, {Tremonti}, Christy, {Brinchmann}, Jarle, {Bruzual}, Gustavo, {Peng}, Eric W., {Seibert}, Mark, {Bernardi}, Mariangela, {Blanton}, Michael, {Brinkmann}, Jon, {Castander}, Francisco, {Cs{\'a}bai}, Istvan, {Fukugita}, Masataka, {Ivezic}, Zeljko, {Munn}, Jeffrey A., {Nichol}, Robert C., {Padmanabhan}, Nikhil, {Thakar}, Aniruddha R., {Weinberg}, David H., \&{York}, Donald
}]{Kauffmann03}
{Kauffmann G.}, {et~al.} 2003, \mnras, 341, 33

\bibitem[{
{Kawata} \& {Mulchaey}(2008)
{Kawata}, Daisuke, \&{Mulchaey}, John S.
}]{Kawata08}
{Kawata D.} \& {Mulchaey J.~S.} 2008, \apjl, 672, L103
 
\bibitem[{
{Kawinwanichakij}{ et al.}(2017)
{Kawinwanichakij}, Lalitwadee,{Papovich}, Casey,{Quadri}, Ryan F.,{Glazebrook}, Karl,{Kacprzak}, Glenn G.,{Allen}, Rebecca J.,{Bell}, Eric F.,{Croton}, Darren J.,{Dekel}, Avishai,{Ferguson}, Henry C.,{Forrest}, Ben,{Grogin}, Norman A.,{Guo}, Yicheng,{Kocevski}, Dale D.,{Koekemoer}, Anton M.,{Labb{\'e}}, Ivo,{Lucas}, Ray A.,{Nanayakkara}, Themiya,{Spitler}, Lee R.,{Straatman}, Caroline M.~S.,{Tran}, Kim-Vy H.,{Tomczak}, Adam,\&{van Dokkum}, Pieter
}]{Kawinwanichakij17}
{Kawinwanichakij L.}, {et~al.} 2017, \apj, 847, 134

\bibitem[{
{Kitaura}{ et al.}(2021)
{Kitaura},{Francisco-Shu},{Ata},{Metin},{Rodr{\'\i}guez-Torres},{Sergio A.},{Hern{\'a}ndez-S{\'a}nchez},{M{\'o}nica},{Balaguera-Antol{\'\i}nez},{A.},\&{Yepes},{Gustavo}
}]{Kitaura21}
{Kitaura F.-S.}, {et~al.} 2021, \mnras, 502, 3456

\bibitem[{
{Kriek}{ et al.}(2015)
{Kriek}, Mariska, {Shapley}, Alice E., {Reddy}, Naveen A., {Siana}, Brian, {Coil}, Alison L., {Mobasher}, Bahram, {Freeman}, William R., {de Groot}, Laura, {Price}, Sedona H., {Sanders}, Ryan, {Shivaei}, Irene, {Brammer}, Gabriel B., {Momcheva}, Ivelina G., {Skelton}, Rosalind E., {van Dokkum}, Pieter G., {Whitaker}, Katherine E., {Aird}, James, {Azadi}, Mojegan, {Kassis}, Marc, {Bullock}, James S., {Conroy}, Charlie, {Dav{\'e}}, Romeel, {Kere{\v{s}}}, Du{\v{s}}an, \&{Krumholz}, Mark
}]{Kriek15}
{Kriek M.}, {et~al.} 2015, \apjs, 218, 15

\bibitem[{
{Krishnan}{ et al.}(2017)
{Krishnan},{Charutha},{Hatch},{Nina A.},{Almaini},{Omar},{Kocevski},{Dale},{Cooke},{Elizabeth A.},{Hartley},{William G.},{Hasinger},{Guenther},{Maltby},{David T.},{Muldrew},{Stuart I.},\&{Simpson},{Chris}
}]{Krishnan17}
{Krishnan C.}, {et~al.} 2017, \mnras, 470, 2170

\bibitem[{
{Krolewski}{ et al.}(2017)
{Krolewski}, Alex, {Lee}, Khee-Gan, {Luki{\'c}}, Zarija, \&{White}, Martin
}]{Krolewski17}
{Krolewski A.}, {et~al.} 2017, \apj, 837, 31

\bibitem[{
{Krolewski}{ et al.}(2018)
{Krolewski},{Alex},{Lee},{Khee-Gan},{White},{Martin},{Hennawi},{Joseph F.},{Schlegel},{David J.},{Nugent},{Peter E.},{Luki{\'c}},{Zarija},{Stark},{Casey W.},{Koekemoer},{Anton M.},{Le F{\`e}vre},{Olivier},{Lemaux},{Brian C.},{Maier},{Christian},{Rich},{R. Michael},{Salvato},{Mara},\&{Tasca},{Lidia}
}]{Krolewski18}
{Krolewski A.}, {et~al.} 2018, \apj, 861, 60

\bibitem[{
{Larson}{ et al.}(1980)
{Larson}, R.~B., {Tinsley}, B.~M., \&{Caldwell}, C.~N.
}]{Larson80}
{Larson R.~B.}, {Tinsley B.~M.} \& {Caldwell C.~N.} 1980, \apj, 237, 692

\bibitem[{
{Le F{\`e}vre}{ et al.}(2015)
{Le F{\`e}vre}, O., {Tasca}, L.~A.~M., {Cassata}, P., {Garilli}, B., {Le Brun}, V., {Maccagni}, D., {Pentericci}, L., {Thomas}, R., {Vanzella}, E., {Zamorani}, G., {Zucca}, E., {Amorin}, R., {Bardelli}, S., {Capak}, P., {Cassar{\`a}}, L., {Castellano}, M., {Cimatti}, A., {Cuby}, J.~G., {Cucciati}, O., {de la Torre}, S., {Durkalec}, A., {Fontana}, A., {Giavalisco}, M., {Grazian}, A., {Hathi}, N.~P., {Ilbert}, O., {Lemaux}, B.~C., {Moreau}, C., {Paltani}, S., {Ribeiro}, B., {Salvato}, M., {Schaerer}, D., {Scodeggio}, M., {Sommariva}, V., {Talia}, M., {Taniguchi}, Y., {Tresse}, L., {Vergani}, D., {Wang}, P.~W., {Charlot}, S., {Contini}, T., {Fotopoulou}, S., {L{\'o}pez-Sanjuan}, C., {Mellier}, Y., \&{Scoville}, N.
}]{LeFevre15}
{Le F{\`e}vre O.}, {et~al.} 2015, \aap, 576, 79

\bibitem[{
{Lee}{ et al.}(2018)
{Lee}, Khee-Gan, {Krolewski}, Alex, {White}, Martin, {Schlegel}, David, {Nugent}, Peter E., {Hennawi}, Joseph F., {M{\"u}ller}, Thomas, {Pan}, Richard, {Prochaska}, J. Xavier, {Font-Ribera}, Andreu, {Suzuki}, Nao, {Glazebrook}, Karl, {Kacprzak}, Glenn G., {Kartaltepe}, Jeyhan S., {Koekemoer}, Anton M., {Le F{\`e}vre}, Olivier, {Lemaux}, Brian C., {Maier}, Christian, {Nanayakkara}, Themiya, {Rich}, R. Michael, {Sanders}, D.~B., {Salvato}, Mara, {Tasca}, Lidia, \&{Tran}, Kim-Vy H.
}]{Lee18}
{Lee K.-G.}, {et~al.} 2018, \apjs, 237, 31

\bibitem[{
{Lee}{ et al.}(2016)
{Lee}, Khee-Gan, {Hennawi}, Joseph F., {White}, Martin, {Prochaska}, J. Xavier, {Font-Ribera}, Andreu, {Schlegel}, David J., {Rich}, R. Michael, {Suzuki}, Nao, {Stark}, Casey W., {Le F{\`e}vre}, Olivier, {Nugent}, Peter E., {Salvato}, Mara, \&{Zamorani}, Gianni
}]{Lee16}
{Lee K.-G.}, {et~al.} 2016, \apj, 817, 160

\bibitem[{
{Lee}{ et al.}(2014)
{Lee},K.-G.,{Hennawi},J.~F.,{Stark},C.,{Prochaska},J.~X.,{White},M.,{Schlegel},D.~J.,{Eilers},A.-C.,{Arinyo-i-Prats},A.,{Suzuki},N.,{Croft},R.~A.~C.,{Caputi},K.~I.,{Cassata},P.,{Ilbert},O.,{Garilli},B.,{Koekemoer},A.~M.,{Le Brun},V.,{Le F{\`e}vre},O.,{Maccagni},D.,{Nugent},P.,{Taniguchi},Y.,{Tasca},L.~A.~M.,{Tresse},L.,{Zamorani},G.,\&{Zucca},E.
}]{Lee14}
{Lee K.-G.}, {et~al.} 2014, \apjl, 795, L12

\bibitem[{
{Lehmer}{ et al.}(2009)
{Lehmer}, B.~D.,{Alexander}, D.~M.,{Geach}, J.~E.,{Smail}, Ian,{Basu-Zych}, A.,{Bauer}, F.~E.,{Chapman}, S.~C.,{Matsuda}, Y.,{Scharf}, C.~A.,{Volonteri}, M.,\&{Yamada}, T.
}]{Lehmer09}
{Lehmer B.~D.}, {et~al.} 2009, \apj, 691, 687

\bibitem[{
{Lemaux}{ et al.}(2022)
{Lemaux},B.~C.,{Cucciati}, O.,{Le F{\`e}vre}, O.,{Zamorani}, G.,{Lubin}, L.~M.,{Hathi}, N.,{Ilbert}, O.,{Pelliccia}, D.,{Amor{\'\i}n}, R.,{Bardelli}, S.,{Cassata}, P.,{Gal}, R.~R.,{Garilli}, B.,{Guaita}, L.,{Giavalisco}, M.,{Hung}, D.,{Koekemoer}, A.,{Maccagni}, D.,{Pentericci}, L.,{Ribeiro}, B.,{Schaerer}, D.,{Shah}, E.,{Shen}, L.,{Staab}, P.,{Talia}, M.,{Thomas}, R.,{Tomczak}, A.~R.,{Tresse}, L.,{Vanzella}, E.,{Vergani}, D.,\&{Zucca}, E.
}]{Lemaux22}
{Lemaux B.~C.}, {et~al.} 2022, \aap, 662, 33

\bibitem[{
{Liang}{ et al.}(2021)
{Liang},{Yongming},{Kashikawa},{Nobunari},{Cai},{Zheng},{Fan},{Xiaohui},{Prochaska},{J. Xavier},{Shimasaku},{Kazuhiro},{Tanaka},{Masayuki},{Uchiyama},{Hisakazu},{Ito},{Kei},{Shimakawa},{Rhythm},{Nagamine},{Kentaro},{Shimizu},{Ikkoh},{Onoue},{Masafusa},\&{Toshikawa},{Jun}
}]{Liang21}
{Liang Y.}, {et~al.} 2021, \apj, 907, 3

\bibitem[{
{Lilly}{ et al.}(2007)
{Lilly}, S.~J., {Le F{\`e}vre}, O., {Renzini}, A., {Zamorani}, G., {Scodeggio}, M., {Contini}, T., {Carollo}, C.~M., {Hasinger}, G., {Kneib}, J. -P., {Iovino}, A., {Le Brun}, V., {Maier}, C., {Mainieri}, V., {Mignoli}, M., {Silverman}, J., {Tasca}, L.~A.~M., {Bolzonella}, M., {Bongiorno}, A., {Bottini}, D., {Capak}, P., {Caputi}, K., {Cimatti}, A., {Cucciati}, O., {Daddi}, E., {Feldmann}, R., {Franzetti}, P., {Garilli}, B., {Guzzo}, L., {Ilbert}, O., {Kampczyk}, P., {Kovac}, K., {Lamareille}, F., {Leauthaud}, A., {Le Borgne}, J. -F., {McCracken}, H.~J., {Marinoni}, C., {Pello}, R., {Ricciardelli}, E., {Scarlata}, C., {Vergani}, D., {Sanders}, D.~B., {Schinnerer}, E., {Scoville}, N., {Taniguchi}, Y., {Arnouts}, S., {Aussel}, H., {Bardelli}, S., {Brusa}, M., {Cappi}, A., {Ciliegi}, P., {Finoguenov}, A., {Foucaud}, S., {Franceschini}, A., {Halliday}, C., {Impey}, C., {Knobel}, C., {Koekemoer}, A., {Kurk}, J., {Maccagni}, D., {Maddox}, S., {Marano}, B., {Marconi}, G., {Meneux}, B., {Mobasher}, B., {Moreau}, C., {Peacock}, J.~A., {Porciani}, C., {Pozzetti}, L., {Scaramella}, R., {Schiminovich}, D., {Shopbell}, P., {Smail}, I., {Thompson}, D., {Tresse}, L., {Vettolani}, G., {Zanichelli}, A., \&{Zucca}, E.
}]{Lilly07}
{Lilly S.~J.}, {et~al.} 2007, \apjs, 172, 70

\bibitem[{
{Lin}{ et al.}(2016)
{Lin},Lihwai,{Capak}, P.~L.,{Laigle}, C.,{Ilbert}, O.,{Hsieh}, Bau-Ching,{Jian}, Hung-Yu,{Lemaux}, B.~C.,{Silverman}, J.~D.,{Coupon}, Jean,{McCracken}, H.~J.,{Hasinger}, G.,{Le F{\'e}vre}, O.,\&{Scoville}, N.
}]{Lin16}
{Lin L.}, {et~al.} 2016, \apj, 817, 97

\bibitem[{
{Lopes}{ et al.}(2017)
{Lopes},P.~A.~A.,{Ribeiro}, A.~L.~B.,\&{Rembold}, S.~B.
}]{Lopes17}
{Lopes P.~A.~A.}, {Ribeiro A.~L.~B.} \& {Rembold S.~B.} 2017, \mnras, 472, 409

\bibitem[{
{Lotz}{ et al.}(2013)
{Lotz},{Jennifer M.},{Papovich},{Casey},{Faber},{S.~M.},{Ferguson},{Henry C.},{Grogin},{Norman},{Guo},{Yicheng},{Kocevski},{Dale},{Koekemoer},{Anton M.},{Lee},{Kyoung-Soo},{McIntosh},{Daniel},{Momcheva},{Ivelina},{Rudnick},{Gregory},{Saintonge},{Amelie},{Tran},{Kim-Vy},{van der Wel},{Arjen},\&{Willmer},{Christopher}
}]{Lotz13}
{Lotz J.~M.}, {et~al.} 2013, \apj, 773, 154

\bibitem[{
{Macuga}{ et al.}(2019)
{Macuga},{Michael},{Martini},{Paul},{Miller},{Eric D.},{Brodwin},{Mark},{Hayashi},{Masao},{Kodama},{Tadayuki},{Koyama},{Yusei},{Overzier},{Roderik A.},{Shimakawa},{Rhythm},{Tadaki},{Ken-ichi},\&{Tanaka},{Ichi}
}]{Macuga19}
{Macuga M.}, {et~al.} 2019, \apj, 874, 54

\bibitem[{
{Man}{ et al.}(2019)
{Man},{Zhong-yi},{Peng},{Ying-jie},{Kong},{Xu},{Guo},{Ke-xin},{Zhang},{Cheng-peng},\&{Dou},{Jing}
}]{Man19}
{Man Z.}, {et~al.} 2019, \mnras, 488, 89

\bibitem[{
{Martini}{ et al.}(2013)
{Martini},Paul,{Miller}, E.~D.,{Brodwin}, M.,{Stanford}, S.~A.,{Gonzalez}, Anthony H.,{Bautz}, M.,{Hickox}, R.~C.,{Stern}, D.,{Eisenhardt}, P.~R.,{Galametz}, A.,{Norman}, D.,{Jannuzi}, B.~T.,{Dey}, A.,{Murray}, S.,{Jones}, C.,\&{Brown}, M.~J.~I.
}]{Martini13}
{Martini P.}, {et~al.} 2013, \apj, 768, 1

\bibitem[{
{McCarthy}{ et al.}(2008)
{McCarthy}, I.~G., {Frenk}, C.~S., {Font}, A.~S., {Lacey}, C.~G., {Bower}, R.~G., {Mitchell}, N.~L., {Balogh}, M.~L., \&{Theuns}, T.
}]{McCarthy08}
{McCarthy I.~G.}, {et~al.} 2008, \mnras, 383, 593


\bibitem[{
{McNab}{ et al.}(2021)
{McNab}, Karen, {Balogh}, Michael L., {van der Burg}, Remco F.~J., {Forestell}, Anya, {Webb}, Kristi, {Vulcani}, Benedetta, {Rudnick}, Gregory, {Muzzin}, Adam, {Cooper}, M.~C., {McGee}, Sean, {Biviano}, Andrea, {Cerulo}, Pierluigi, {Chan}, Jeffrey C.~C., {De Lucia}, Gabriella, {Demarco}, Ricardo, {Finoguenov}, Alexis, {Forrest}, Ben, {Golledge}, Caelan, {Jablonka}, Pascale, {Lidman}, Chris, {Nantais}, Julie, {Old}, Lyndsay, {Pintos-Castro}, Irene, {Poggianti}, Bianca, {Reeves}, Andrew M.~M., {Wilson}, Gillian, {Yee}, Howard K.~C., \&{Zaritsky}, Dennis
}]{McNab21}
{McNab K.}, {et~al.} 2021, \mnras, 508, 157

\bibitem[{
{Micha{\l}owski}{ et al.}(2017)
{Micha{\l}owski}, Micha{\l} J., {Dunlop}, J.~S., {Koprowski}, M.~P., {Cirasuolo}, M., {Geach}, J.~E., {Bowler}, R.~A.~A., {Mortlock}, A., {Caputi}, K.~I., {Aretxaga}, I., {Arumugam}, V., {Chen}, Chian-Chou, {McLure}, R.~J., {Birkinshaw}, M., {Bourne}, N., {Farrah}, D., {Ibar}, E., {van der Werf}, P., \&{Zemcov}, M.
}]{Michalowski17}
{Micha{\l}owski M.}, {et~al.} 2017, \mnras, 469, 492

\bibitem[{
{Miller}{ et al.}(2015)
{Miller},{Tim B.},{Hayward},{Christopher C.},{Chapman},{Scott C. },\&{Behroozi},{Peter S.}
}]{Miller15}
{Miller T.~B.}, {et~al.} 2015, \mnras, 452, 878

\bibitem[{
{Miller}{ et al.}(2003)
{Miller},{Christopher J.},{Nichol},{Robert C.},{G{\'o}mez},{Percy L.},{Hopkins},{Andrew M.},\&{Bernardi},{Mariangela}
}]{Miller03}
{Miller T.~B.}, {et~al.} 2003, \apj, 597, 142

\bibitem[{
{Miraghaei}(2020)
{Miraghaei},{Halime}
}]{Miraghaei20}
{Miraghaei H.}, 2020, \aj, 160, 227

\bibitem[{
{Momose}{ et al.}(2021b)
{Momose},{Rieko},{Shimasaku},{Kazuhiro},{Nagamine},{Kentaro},{Shimizu},{Ikkoh},{Kashikawa},{Nobunari},{Ando},{Makoto},\&{Kusakabe},{Haruka}
}]{Momose21c}
{Momose R.}, {et~al.} 2021b, \apjl, 912, L24

\bibitem[{
{Momose}{ et al.}(2021a)
{Momose},{Rieko},{Shimasaku},{Kazuhiro},{Kashikawa},{Nobunari},{Nagamine},{Kentaro},{Shimizu},{Ikkoh},{Nakajima},{Kimihiko},{Terao},{Yasunori},{Kusakabe},{Haruka},{Ando},{Makoto},{Motohara},{Kentaro},\&{Spitler},{Lee}
}]{Momose21a}
{Momose R.}, {et~al.} 2021a, \apj, 909, 117

\bibitem[{
{Moore}{ et al.}(1998)
{Moore}, Ben, {Lake}, George, \&{Katz}, Neal
}]{Moore98}
{Moore B.}, {Lake G.} \& {Katz N.} 1998, \apj, 495, 139

\bibitem[{
{Moore}{ et al.}(1996)
{Moore}, Ben, {Katz}, Neal, {Lake}, George, {Dressler}, Alan, {Oemler}, Augustus
}]{Moore96}
{Moore B.}, {et~al.} 1996, \nat, 379, 613

\bibitem[{
{Moster}{ et al.}(2013)
{Moster}, Benjamin P., {Naab}, Thorsten, \& {White}, Simon D.~M.
}]{Moster13}
{Moster B.~P.}, {Naab T.}, \&{White S.~D.~M}, 2013, \mnras, 428, 3121

\bibitem[{
{Muzzin}{ et al.}(2012)
{Muzzin}, Adam, {Wilson}, Gillian, {Yee}, H.~K.~C., {Gilbank}, David, {Hoekstra}, Henk, {Demarco}, Ricardo, {Balogh}, Michael, {van Dokkum}, Pieter, {Franx}, Marijn, {Ellingson}, Erica, {Hicks}, Amalia, {Nantais}, Julie, {Noble}, Allison, {Lacy}, Mark, {Lidman}, Chris, {Rettura}, Alessandro, {Surace}, Jason,\& {Webb}, Tracy
}]{Muzzin12}
{Muzzin A.}, {et~al.} 2012, \apj, 746, 188

\bibitem[{
{Nakajima}{ et al.}(2013)
{Nakajima}, Kimihiko, {Ouchi}, Masami, {Shimasaku}, Kazuhiro, {Hashimoto}, Takuya, {Ono}, Yoshiaki, \&{Lee}, Janice C.
}]{Nakajima13}
{Nakajima K.}, {et~al.} 2013, \apj, 769, 3

\bibitem[{
{Nanayakkara}{ et al.}(2016)
{Nanayakkara}, Themiya, {Glazebrook}, Karl, {Kacprzak}, Glenn G., {Yuan}, Tiantian, {Tran}, Kim-Vy, {Spitler}, Lee, {Kewley}, Lisa, {Straatman}, Caroline, {Cowley}, Michael, {Fisher}, David, {Labbe}, Ivo, {Tomczak}, Adam, {Allen}, Rebecca, \&{Alcorn}, Leo
}]{Nanayakkara16}
{Nanayakkara T.}, {et~al.} 2016, \apj, 828, 21

\bibitem[{
{Nantais}{ et al.}(2016)
{Nantais}, Julie B.,{van der Burg}, Remco F.~J.,{Lidman}, Chris,{Demarco}, Ricardo,{Noble}, Allison,{Wilson}, Gillian,{Muzzin}, Adam,{Foltz}, Ryan,{DeGroot}, Andrew,\&{Cooper}, Michael C.
}]{Nantais16}
{Nantais J.~B.}, {et~al.} 2016, \aap, 592, 161

\bibitem[{
{Overzier}{ et al.}(2008)
{Overzier}, Roderik A.,{Bouwens}, R.~J.,{Cross}, N.~J.~G.,{Venemans}, B.~P.,{Miley}, G.~K.,{Zirm}, A.~W.,{Ben{\'\i}tez}, N.,{Blakeslee}, J.~P.,{Coe}, D.,{Demarco}, R.,{Ford}, H.~C.,{Homeier}, N.~L.,{Illingworth}, G.~D.,{Kurk}, J.~D.,{Martel}, A.~R.,{Mei}, S.,{Oliveira}, I.,{R{\"o}ttgering}, H.~J.~A.,{Tsvetanov}, Z.~I.,\&{Zheng}, W.
}]{Overzier08}
{Overzier R.~A.}, {et~al.} 2008, \apj, 673, 143

\bibitem[{
{Pallero}{ et al.}(2019)
{Pallero}, Diego, {G{\'o}mez}, Facundo A., {Padilla}, Nelson D., {Torres-Flores}, S., {Demarco}, R., {Cerulo}, P., \&{Olave-Rojas}, D.
}]{Pallero19}
{Pallero D.}, {et~al.} 2019, \mnras, 488, 847

\bibitem[{
{Peng}{ et al.}(2012)
{Peng}, Ying-jie, {Lilly}, Simon J., {Renzini}, Alvio,\& {Carollo}, Marcella
}]{Peng12}
{Peng Y.}, {et~al.} 2012, \apj, 757, 4

\bibitem[{
{Peng}{ et al.}(2010)
{Peng}, Ying-jie, {Lilly}, Simon J., {Kova{\v{c}}}, Katarina, {Bolzonella}, Micol, {Pozzetti}, Lucia, {Renzini}, Alvio, {Zamorani}, Gianni, {Ilbert}, Olivier, {Knobel}, Christian, {Iovino}, Angela, {Maier}, Christian, {Cucciati}, Olga, {Tasca}, Lidia, {Carollo}, C. Marcella, {Silverman}, John, {Kampczyk}, Pawel, {de Ravel}, Loic, {Sanders}, David, {Scoville}, Nicholas, {Contini}, Thierry, {Mainieri}, Vincenzo, {Scodeggio}, Marco, {Kneib}, Jean-Paul, {Le F{\`e}vre}, Olivier, {Bardelli}, Sandro, {Bongiorno}, Angela, {Caputi}, Karina, {Coppa}, Graziano, {de la Torre}, Sylvain, {Franzetti}, Paolo, {Garilli}, Bianca, {Lamareille}, Fabrice, {Le Borgne}, Jean-Francois, {Le Brun}, Vincent, {Mignoli}, Marco, {Perez Montero}, Enrique, {Pello}, Roser, {Ricciardelli}, Elena, {Tanaka}, Masayuki, {Tresse}, Laurence, {Vergani}, Daniela, {Welikala}, Niraj, {Zucca}, Elena, {Oesch}, Pascal, {Abbas}, Ummi, {Barnes}, Luke, {Bordoloi}, Rongmon, {Bottini}, Dario, {Cappi}, Alberto, {Cassata}, Paolo, {Cimatti}, Andrea, {Fumana}, Marco, {Hasinger}, Gunther, {Koekemoer}, Anton, {Leauthaud}, Alexei, {Maccagni}, Dario, {Marinoni}, Christian, {McCracken}, Henry, {Memeo}, Pierdomenico, {Meneux}, Baptiste, {Nair}, Preethi, {Porciani}, Cristiano, {Presotto}, Valentina, {Scaramella}, Roberto
}]{Peng10}
{Peng Y.}, {et~al.} 2010, \apj, 721, 193

\bibitem[{
{Pimbblet}{ et al.}(2013)
{Pimbblet},K.~A.,{Shabala},S.~S.,{Haines},C.~P.,{Fraser-McKelvie},A.,\&{Floyd},D.~J.~E.
}]{Pimbblet13}
{Pimbblet K.~A.}, {et~al.} 2013, \mnras, 429, 1827

\bibitem[{
{Polletta}{ et al.}(2021)
{Polletta}, M.,{Soucail}, G.,{Dole}, H.,{Lehnert}, M.~D.,{Pointecouteau}, E.,{Vietri}, G.,{Scodeggio}, M.,{Montier}, L.,{Koyama}, Y.,{Lagache}, G.,{Frye}, B.~L.,{Cusano}, F.,\&{Fumana}, M.
}]{Polletta21}
{Polletta M.}, {et~al.} 2021, \aap, 654, 121

\bibitem[{
{Quadri}{ et al.}(2012)
{Quadri},{Ryan F.},{Williams},{Rik J.},{Franx},{Marijn},\&{Hildebrandt},{Hendrik}
}]{Quadri12}
{Quadri R.~F.}, {et~al.} 2012, \apj, 744, 88

\bibitem[{
{Reeves}{ et al.}(2021)
{Reeves},{Andrew M.~M.},{Balogh},{Michael L.},{van der Burg},{Remco F.~J.},{Finoguenov},{Alexis},{Kukstas},{Egidijus},{McCarthy},{Ian G.},{Webb},{Kristi},{Muzzin},{Adam},{McGee},{Sean},{Rudnick},{Gregory},{Biviano},{Andrea},{Cerulo},{Pierluigi},{Chan},{Jeffrey C.~C.},{Cooper},{M.~C.},{Demarco},{Ricardo},{Jablonka},{Pascale},{De Lucia},{Gabriella},{Vulcani},{Benedetta},{Wilson},{Gillian},{Yee},{Howard K.~C.},\&{Zaritsky},{Dennis}
}]{Reeves21}
{Reeves A.~M.~M.}, {et~al.} 2021, \mnras, 506, 3364

\bibitem[{
{Rodr{\'\i}guez Montero}{ et al.}(2019)
{Rodr{\'\i}guez Montero}, Francisco and {Dav{\'e}}, Romeel and {Wild}, Vivienne and {Angl{\'e}s-Alc{\'a}zar}, Daniel, \&{Narayanan}, Desika
}]{Rodriguez19}
{Rodr{\'\i}guez Montero F.}, {et~al.} 2019, \mnras, 490, 2139

\bibitem[{
{Sabater}{ et al.}(2019)
{Sabater}, J.,{Best}, P.~N.,{Hardcastle}, M.~J.,{Shimwell}, T.~W.,{Tasse}, C.,{Williams}, W.~L.,{Br{\"u}ggen}, M.,{Cochrane}, R.~K.,{Croston}, J.~H.,{de Gasperin}, F.,{Duncan}, K.~J.,{G{\"u}rkan}, G.,{Mechev}, A.~P.,{Morabito}, L.~K.,{Prandoni}, I.,{R{\"o}ttgering}, H.~J.~A.,{Smith}, D.~J.~B.,{Harwood}, J.~J.,{Mingo}, B.,{Mooney}, S.,\&{Saxena}, A.
}]{Sabater19}
{Sabater J.}, {et~al.} 2019, \aap, 622, 17

\bibitem[{
{Sanders} \& {Mirabel}(1996)
{Sanders}, D.~B., \&{Mirabel}, I.~F.
}]{Sanders96}
{Sanders D.~B.} \& {Mirabel I.~F.} 1996, \araa, 34, 749

\bibitem[{
{Schawinski}{ et al.}(2009)
{Schawinski}, Kevin, {Lintott}, Chris J., {Thomas}, Daniel, {Kaviraj}, Sugata, {Viti}, Serena, {Silk}, Joseph, {Maraston}, Claudia, {Sarzi}, Marc, {Yi}, Sukyoung K., {Joo}, Seok-Joo, {Daddi}, Emanuele, {Bayet}, Estelle, {Bell}, Tom, \&{Zuntz}, Joe
}]{Schawinski09}
{Schawinski K.}, {et ~al.} 2009, \apj, 690, 1672

\bibitem[{
{Scoville}{ et al.}(2007)
{Scoville},N.,{Aussel},H.,{Brusa},M.,{Capak},P.,{Carollo},C.~M.,{Elvis},M.,{Giavalisco},M.,{Guzzo},L.,{Hasinger},G.,{Impey},C.,{Kneib},J.-P.,{LeFevre},O.,{Lilly},S.~J.,{Mobasher},B.,{Renzini},A.,{Rich},R.~M.,{Sanders},D.~B.,{Schinnerer},E.,{Schminovich},D.,{Shopbell},P.,{Taniguchi},Y.,\&{Tyson},N.~D.
}]{Scoville07}
{Scoville N.}, {et al.} 2007, \apjs, 172, 1

\bibitem[{
{Shah}{ et al.}(2020)
{Shah}, Ekta A., {Kartaltepe}, Jeyhan S., {Magagnoli}, Christina T., {Cox}, Isabella G., {Wetherell}, Caleb T., {Vanderhoof}, Brittany N., {Calabro}, Antonello, {Chartab}, Nima, {Conselice}, Christopher J., {Croton}, Darren J., {Donley}, Jennifer, {de Groot}, Laura, {de la Vega}, Alexander, {Hathi}, Nimish P., {Ilbert}, Olivier, {Inami}, Hanae, {Kocevski}, Dale D., {Koekemoer}, Anton M., {Lemaux}, Brian C., {Mantha}, Kameswara Bharadwaj, {Marchesi}, Stefano, {Martig}, Marie, {Masters}, Daniel C., {McGrath}, Elizabeth J., {McIntosh}, Daniel H., {Moreno}, Jorge, {Nayyeri}, Hooshang, {Pampliega}, Belen Alcalde, {Salvato}, Mara, {Snyder}, Gregory F., {Straughn}, Amber N., {Treister}, Ezequiel, \&{Weston}, Madalyn E.
}]{Shah20}
{Shah E.~A.}, {et al.} 2020, \apj, 904, 107

\bibitem[{
{Shi}{ et al.}(2019)
{Shi},{Ke},{Huang},{Yun},{Lee},{Kyoung-Soo},{Toshikawa},{Jun},{Bowen},{Kathryn N.},{Malavasi},{Nicola},{Lemaux},{B.~C.},{Cucciati},{Olga},{Le Fevre},{Olivier},\&{Dey},{Arjun}
}]{Shi19}
{Shi K.}, {et al.} 2019, \apjs, 879, 9

\bibitem[{
{Shibuya}{ et al.}(2014)
{Shibuya}, Takatoshi, {Ouchi}, Masami, {Nakajima}, Kimihiko, {Hashimoto}, Takuya, {Ono}, Yoshiaki, {Rauch}, Michael, {Gauthier}, Jean-Rene, {Shimasaku}, Kazuhiro, {Goto}, Ryosuke, {Mori}, Masao, \&{Umemura.}, Masayuki
}]{Shibuya14}
{Shibuya T.}, {et al.} 2014, \apj, 788, 74

\bibitem[{
{Shimakawa}{ et al.}(2017)
{Shimakawa},{Rhythm},{Kodama},{Tadayuki},{Hayashi},{Masao},{Tanaka},{Ichi},{Matsuda},{Yuichi},{Kashikawa},{Nobunari},{Shibuya},{Takatoshi},{Tadaki},{Ken-ichi},{Koyama},{Yusei},{Suzuki},{Tomoko L.},\&{Yamamoto},{Moegi}
}]{Shimakawa17}
{Shimakawa R.}, {et al.} 2017, \mnras, 468, 21

\bibitem[{
{Silva}{ et al.}(2018)
{Silva}, Andrea, {Marchesini}, Danilo, {Silverman}, John D., {Skelton}, Rosalind, {Iono}, Daisuke, {Martis}, Nicholas, {Marsan}, Z. Cemile, {Tadaki}, Ken-ichi, {Brammer}, Gabriel, \&{kartaltepe}, Jeyhan
}]{Silva18}
{Silva A.}, {et al.} 2018, \apj, 868, 46

\bibitem[{
{Silverman}{ et al.}(2009)
{Silverman}, J.~D.,{Kova{\v{c}}}, K.,{Knobel}, C.,{Lilly}, S.,{Bolzonella}, M.,{Lamareille}, F.,{Mainieri}, V.,{Brusa}, M.,{Cappelluti}, N.,{Peng}, Y.,{Hasinger}, G.,{Zamorani}, G.,{Scodeggio}, M.,{Contini}, T.,{Carollo}, C.~M.,{Jahnke}, K.,{Kneib}, J.-P.,{Le Fevre}, O.,{Bardelli}, S.,{Bongiorno}, A.,{Brunner}, H.,{Caputi}, K.,{Civano}, F.,{Comastri}, A.,{Coppa}, G.,{Cucciati}, O.,{de la Torre}, S.,{de Ravel}, L. ,{Elvis}, M.,{Finoguenov}, A.,{Fiore}, F.,{Franzetti}, P.,{Garilli}, B.,{Gilli}, R.,{Griffiths}, R.,{Iovino}, A.,{Kampczyk}, P.,{Koekemoer}, A.,{Le Borgne}, J.-F.,{Le Brun}, V.,{Maier}, C.,{Mignoli}, M.,{Pello}, R.,{Perez Montero}, E.,{Ricciardelli}, E.,{Tanaka}, M.,{Tasca}, L.,{Tresse}, L.,{Vergani}, D.,{Vignali}, C.,{Zucca}, E.,{Bottini}, D.,{Cappi}, A.,{Cassata}, P.,{Marinoni}, C.,{McCracken}, H.~J. ,{Memeo}, P.,{Meneux}, B.,{Oesch}, P.,{Porciani}, C.,\&{Salvato}, M.
}]{Silverman09}
{Silverman J.~D.}, {et al.} 2009, \apj, 695, 171

\bibitem[{
{Smith}{ et al.}(2016)
{Smith}, Rory, {Choi}, Hoseung, {Lee}, Jaehyun, {Rhee}, Jinsu, {Sanchez-Janssen}, Ruben, \&{Yi}, Sukyoung K.
}]{Smith16}
{Smith K.}, {et al.} 2016, \apj, 833, 109

\bibitem[{
{Smith}{ et al.}(2015)
{Smith}, R., {S{\'a}nchez-Janssen}, R., {Beasley}, M.~A., {Candlish}, G.~N., {Gibson}, B.~K., {Puzia}, T.~H., {Janz}, J., {Knebe}, A., {Aguerri}, J.~A.~L., {Lisker}, T., {Hensler}, G., {Fellhauer}, M., {Ferrarese}, L.,\& {Yi}, S.~K.
}]{Smith15}
{Smith R.}, {et al.} 2015, \mnras, 454, 2502

\bibitem[{
{Smith}{ et al.}(2013)
{Smith}, R., {S{\'a}nchez-Janssen}, R., {Fellhauer}, M., {Puzia}, T.~H., {Aguerri}, J.~A.~L.,\& {Farias}, J.~P.
}]{Smith13}
{Smith R.}, {et al.} 2013, \mnras, 429, 1066

\bibitem[{
{Smith}{ et al.}(2010)
{Smith}, R., {Davies}, J.~I.,\& {Nelson}, A.~H.
}]{Smith10}
{Smith R.}, {Davies J.~I}, \& {Nelson A.~H.} 2010, \mnras, 405, 1723

\bibitem[{
{Smol{\v{c}}i{\'c}}{ et al.}(2012)
{Smol{\v{c}}i{\'c}}, V., {Aravena}, M., {Navarrete}, F., {Schinnerer}, E., {Riechers}, D.~A., {Bertoldi}, F., {Feruglio}, C., {Finoguenov}, A., {Salvato}, M., {Sargent}, M., {McCracken}, H.~J., {Albrecht}, M., {Karim}, A., {Capak}, P., {Carilli}, C.~L., {Cappelluti}, N., {Elvis}, M., {Ilbert}, O., {Kartaltepe}, J., {Lilly}, S., {Sanders}, D., {Sheth}, K., {Scoville}, N.~Z., \&{Taniguchi}, Y.
}]{Smolcic12}
{Smol{\v{c}}i{\'c} V.}, {et al.} 2012, \aap, 548, 4

\bibitem[{
{Sobral}{ et al.}(2011)
{Sobral}, David, {Best}, Philip N., {Smail}, Ian, {Geach}, James E., {Cirasuolo}, Michele, {Garn}, Timothy,\& {Dalton}, Gavin B.
}]{Sobral11}
{Sobral D.}, {et al.} 2011, \mnras, 411, 675

\bibitem[{
{Speagle}{ et al.}(2014)
{Speagle},{J.~S.},{Steinhardt},{C.~L.},{Capak},{P.~L.},\&{Silverman},{J.~D.}
}]{Speagle14}
{Speagle J.~S.}, {et al.} 2014, \apjs, 214, 15

\bibitem[{
{Springel}{ et al.}(2005)
{Springel}, Volker, {Di Matteo}, Tiziana, \&{Hernquist}, Lars
}]{Springel05a}
{Springel V.}, {Di Matteo T.}, \& {Hernquist L.} 2005, \apjl, 620, L79

\bibitem[{
{Springel}{ et al.}(2005)
{Springel}, Volker, {Di Matteo}, Tiziana, \&{Hernquist}, Lars
}]{Springel05b}
{Springel V.}, {Di Matteo T.}, \& {Hernquist L.} 2005, \mnras, 361, 776

\bibitem[{
{Straatman}{ et al.}(2016)
{Straatman},{Caroline M.~S.},{Spitler},{Lee R.},{Quadri},{Ryan F.},{Labb{\'e}},{Ivo},{Glazebrook},{Karl},{Persson},{S. Eric},{Papovich},{Casey},{Tran},{Kim-Vy H.},{Brammer},{Gabriel B.},{Cowley},{Michael},{Tomczak},{Adam},{Nanayakkara},{Themiya},{Alcorn},{Leo},{Allen},{Rebecca},{Broussard},{Adam},{van Dokkum},{Pieter},{Forrest},{Ben},{van Houdt},{Josha},{Kacprzak},{Glenn G.},{Kawinwanichakij},{Lalitwadee},{Kelson},{Daniel D.},{Lee},{Janice},{McCarthy},{Patrick J.},{Mehrtens},{Nicola},{Monson},{Andrew},{Murphy},{David},{Rees},{Glen},{Tilvi},{Vithal},\&{Whitaker},{Katherine E.}
}]{Straatman16}
{Straatman C.~M.~S.}, {et al.} 2016, \apj, 830, 51

\bibitem[{
{Strateva}{ et al.}(2001)
{Strateva}, Iskra, {Ivezi{\'c}}, {\v{Z}}eljko, {Knapp}, Gillian R., {Narayanan}, Vijay K., {Strauss}, Michael A., {Gunn}, James E., {Lupton}, Robert H., {Schlegel}, David, {Bahcall}, Neta A., {Brinkmann}, Jon, {Brunner}, Robert J., {Budav{\'a}ri}, Tam{\'a}s, {Csabai}, Istv{\'a}n, {Castander}, Francisco Javier, {Doi}, Mamoru, {Fukugita}, Masataka, {Gy{\H{o}}ry}, Zsuzsanna, {Hamabe}, Masaru, {Hennessy}, Greg, {Ichikawa}, Takashi, {Kunszt}, Peter Z., {Lamb}, Don Q., {McKay}, Timothy A., {Okamura}, Sadanori, {Racusin}, Judith, {Sekiguchi}, Maki, {Schneider}, Donald P., {Shimasaku}, Kazuhiro, \&{York}, Donald
}]{Strateva01}
{Strateva I.}, {et al.} 2001, \aj, 122, 1861

\bibitem[{
{Sugai}{ et al.}(2015)
{Sugai},{Hajime},{Tamura},{Naoyuki},{Karoji},{Hiroshi},{Shimono},{Atsushi},{Takato},{Naruhisa},{Kimura},{Masahiko},{Ohyama},{Youichi},{Ueda},{Akitoshi},{Aghazarian},{Hrand},{de Arruda},{Marcio Vital},{Barkhouser},{Robert H.},{Bennett},{Charles L.},{Bickerton},{Steve},{Bozier},{Alexandre},{Braun},{David F.},{Bui},{Khanh},{Capocasale},{Christopher M.},{Carr},{Michael A.},{Castilho},{Bruno},{Chang},{Yin-Chang},{Chen},{Hsin-Yo},{Chou},{Richard C.~Y.},{Dawson},{Olivia R.},{Dekany},{Richard G.},{Ek},{Eric M.},{Ellis},{Richard S.},{English},{Robin J.},{Ferrand},{Didier},{Ferreira},{D{\'e}cio},{Fisher},{Charles D.},{Golebiowski},{Mirek},{Gunn},{James E.},{Hart},{Murdock},{Heckman},{Timothy M.},{Ho},{Paul T.~P.},{Hope},{Stephen},{Hovland},{Larry E.},{Hsu},{Shu-Fu},{Hu},{Yen-Shan},{Huang},{Pin Jie},{Jaquet},{Marc},{Karr},{Jennifer E.},{Kempenaar},{Jason G.},{King},{Matthew E.},{le F{\`e}vre},{Olivier},{Mignant},{David Le},{Ling},{Hung-Hsu},{Loomis},{Craig},{Lupton},{Robert H.},{Madec},{Fabrice},{Mao},{Peter},{Souza Marrara},{Lucas},{M{\'e}nard},{Brice},{Morantz},{Chaz},{Murayama},{Hitoshi},{Murray},{Graham J.},{Cesar de Oliveira},{Antonio},{Mendes de Oliveira},{Claudia},{Souza de Oliveira},{Ligia},{Orndorff},{Joe D.},{de Paiva Vila{\c{c}}a},{Rodrigo},{Partos},{Eamon J.},{Pascal},{Sandrine},{Pegot-Ogier},{Thomas},{Reiley},{Daniel J.},{Riddle},{Reed},{Santos},{Leandro},{dos Santos},{Jesulino Bispo},{Schwochert},{Mark A.},{Seiffert},{Michael D.},{Smee},{Stephen A.},{Smith},{Roger M.},{Steinkraus},{Ronald E.},{Sodr{\'e}},{Laerte},{Spergel},{David N.},{Surace},{Christian},{Tresse},{Laurence},{Vidal},{Cl{\'e}ment},{Vives},{Sebastien},{Wang},{Shiang-Yu},{Wen},{Chih-Yi},{Wu},{Amy C.},{Wyse},{Rosie},\&{Yan},{Chi-Hung}
}]{Sugai15}
{Sugai H.}, {et al.} 2015, Journal of Astronomical Telescopes, Instruments, and Systems, 1, 5001

\bibitem[{
{Tanaka}(2012)
{Tanaka},{Masayuki}
}]{Tanaka12}
{Tanaka M.}, 2012, \pasj, 64, 37

\bibitem[{
{Taranu}{ et al.}(2014)
{Taranu}, Dan S., {Hudson}, Michael J., {Balogh}, Michael L., {Smith}, Russell J., {Power}, Chris, {Oman}, Kyle A., \&{Krane}, Brad
}]{Taranu14}
{Taranu D.~S.}, {et al.} 2014, \mnras, 440, 1934

\bibitem[{
{Tomczak}{ et al.}(2016)
{Tomczak},{Adam R.},{Quadri},{Ryan F.},{Tran},{Kim-Vy H.},{Labb{\'e}},{Ivo},{Straatman},{Caroline M.~S.},{Papovich},{Casey},{Glazebrook},{Karl},{Allen},{Rebecca},{Brammer},{Gabreil B.},{Cowley},{Michael},{Dickinson},{Mark},{Elbaz},{David},{Inami},{Hanae},{Kacprzak},{Glenn G.},{Morrison},{Glenn E.},{Nanayakkara},{Themiya},{Persson},{S. Eric},{Rees},{Glen A.},{Salmon},{Brett},{Schreiber},{Corentin},{Spitler},{Lee R.},\&{Whitaker},{Katherine E.}
}]{Tomczak16}
{Tomczak A.~R.}, {et al.} 2016, \apj, 817, 118

\bibitem[{
{Toomre} \& {Toomre}(1972)
{Toomre}, Alar,\& {Toomre}, Juri
}]{Toomre72}
{Toomre A.} \& {Toomre J.} 1972, \apj, 178, 623

\bibitem[{
{Toshikawa}{ et al.}(2016)
{Toshikawa},{Jun},{Kashikawa},{Nobunari},{Overzier},{Roderik},{Malkan},{Matthew A.},{Furusawa},{Hisanori},{Ishikawa},{Shogo},{Onoue},{Masafusa},{Ota},{Kazuaki},{Tanaka},{Masayuki},{Niino},{Yuu},\&{Uchiyama},{Hisakazu}
}]{Toshikawa16}
{Toshikawa J.}, {et al.} 2016, \apj, 826, 114

\bibitem[{
{Tran}{ et al.}(2010)
{Tran},{Kim-Vy H},{Papovich},{Casey},{Saintonge},{Am{\'e}lie},{Brodwin},{Mark},{Dunlop},{James S.},{Farrah},{Duncan},{Finkelstein},{Keely D.},{Finkelstein},{Steven L.},{Lotz},{Jennifer},{McLure},{Ross J.},{Momcheva},{Ivelina},\&{Willmer},{Christopher N.~A.}
}]{Tran10}
{Tran K.-V.~H.}, {et al.} 2010, \apjl, 719, L126

\bibitem[{
{Umehata}{ et al.}(2015)
{Umehata}, H.,{Tamura}, Y.,{Kohno}, K.,{Ivison}, R.~J.,{Alexander}, D.~M.,{Geach}, J.~E.,{Hatsukade}, B.,{Hughes}, D.~H.,{Ikarashi}, S.,{Kato}, Y.,{Izumi}, T.,{Kawabe}, R.,{Kubo}, M.,{Lee}, M.,{Lehmer}, B.,{Makiya}, R.,{Matsuda}, Y.,{Nakanishi}, K.,{Saito}, T.,{Smail}, I.,{Yamada}, T.,{Yamaguchi}, Y.,\&{Yun}, M.
}]{Umehata15}
{Umehata H.}, {et al.} 2015, \apjl, 815, L8

\bibitem[{
{van de Voort}{ et al.}(2017)
{van de Voort},{Freeke},{Bah{\'e}},{Yannick M.},{Bower},{Richard G.},{Correa},{Camila A.},{Crain},{Robert A.},{Schaye},{Joop},\&{Theuns},{Tom}
}]{vandeVoort17}
{van de Voort F.}, {et al.} 2017, \mnras, 466, 3460

\bibitem[{
{Venemans}{ et al.}(2007)
{Venemans}, B.~P.,{R{\"o}ttgering}, H.~J.~A.,{Miley}, G.~K.,{van Breugel}, W.~J.~M.,{de Breuck}, C.,{Kurk}, J.~D.,{Pentericci}, L.,{Stanford}, S.~A.,{Overzier}, R.~A.,{Croft}, S.,\&{Ford}, H.
}]{Venemans07}
{Venemans B.~P.}, {et al.} 2007, \aap, 461, 823

\bibitem[{
{Vijayaraghavan} \& {Ricker}(2013)
{Vijayaraghavan}, R., \&{Ricker}, P.~M.
}]{Vijayaraghavan13}
{Vijayaraghavan R.}, \& {Ricker P.~M.} 2013, \mnras, 435, 2713

\bibitem[{
{von der Linden}{ et al.}(2010)
{von der Linden},{Anja},{Wild},{Vivienne},{Kauffmann},{Guinevere},{White},{Simon D.~M.},\&{Weinmann},{Simone}
}]{Linden10}
{von der Linden A.}, {et al.} 2010, \mnras, 404, 1231

\bibitem[{
{Watson}{ et al.}(2019)
{Watson},{Courtney},{Tran},{Kim-Vy},{Tomczak},{Adam},{Alcorn},{Leo},{Salazar},{Irene V.},{Gupta},{Anshu},{Momcheva},{Ivelina},{Papovich},{Casey},{van Dokkum},{Pieter},{Brammer},{Gabriel},{Lotz},{Jennifer},\&{Willmer},{Christopher N.~A.}
}]{Watson19}
{Watson C.}, {et al.} 2019, \apj, 874, 63

\bibitem[{
{Weaver}{ et al.}(2023)
{Weaver}, J.~R., {Davidzon}, I., {Toft}, S., {Ilbert}, O., {McCracken}, H.~J., {Gould}, K.~M.~L., {Jespersen}, C.~K., {Steinhardt}, C., {Lagos}, C.~D.~P., {Capak}, P.~L., {Casey}, C.~M., {Chartab}, N., {Faisst}, A.~L., {Hayward}, C.~C., {Kartaltepe}, J.~S., {Kauffmann}, O.~B., {Koekemoer}, A.~M., {Kokorev}, V., {Laigle}, C., {Liu}, D., {Long}, A., {Magdis}, G.~E., {McPartland}, C.~J.~R., {Milvang-Jensen}, B., {Mobasher}, B., {Moneti}, A., {Peng}, Y., {Sanders}, D.~B., {Shuntov}, M., {Sneppen}, A., {Valentino}, F., {Zalesky}, L., \& {Zamorani}, G.
}]{Weaver23}
{Weaver J.~R.}, {et al.} 2023, \aap, 677, 184

\bibitem[{
{Weaver}{ et al.}(2022)
{Weaver}, J.~R., {Kauffmann}, O.~B., {Ilbert}, O., {McCracken}, H.~J., {Moneti}, A., {Toft}, S., {Brammer}, G., {Shuntov}, M., {Davidzon}, I., {Hsieh}, B.~C., {Laigle}, C., {Anastasiou}, A., {Jespersen}, C.~K., {Vinther}, J., {Capak}, P., {Casey}, C.~M., {McPartland}, C.~J.~R., {Milvang-Jensen}, B., {Mobasher}, B., {Sanders}, D.~B., {Zalesky}, L., {Arnouts}, S., {Aussel}, H., {Dunlop}, J.~S., {Faisst}, A., {Franx}, M., {Furtak}, L.~J., {Fynbo}, J.~P.~U., {Gould}, K.~M.~L., {Greve}, T.~R., {Gwyn}, S., {Kartaltepe}, J.~S., {Kashino}, D., {Koekemoer}, A.~M., {Kokorev}, V., {Le Fevre}, O., {Lilly}, S., {Masters}, D., {Magdis}, G., {Mehta}, V., {Peng}, Y., {Riechers}, D.~A., {Salvato}, M., {Sawicki}, M., {Scarlata}, C., {Scoville}, N., {Shirley}, R., {Sneppen}, A., {Smolcic}, V., {Steinhardt}, C., {Stern}, D., {Tanaka}, M., {Taniguchi}, Y., {Teplitz}, H.~I., {Vaccari}, M., {Wang}, W. -H., \&{Zamorani}, G.
}]{Weaver21}
{Weaver J.~R.}, {et al.} 2022, \apjs, 258, 30

\bibitem[{
{Wetzel}{et al.}(2012)
{Wetzel}, Andrew R., {Tinker}, Jeremy L., \&{Conroy}, Charlie
}]{Wetzel12}
{Wetzel A.~R.}, {Tinker J.~L.}, \& {Conroy C.} 2012, \mnras, 424, 232

\bibitem[{
{Whitaker}{ et al.}(2014)
{Whitaker},{Katherine E.},{Franx},{Marijn},{Leja},{Joel},{van Dokkum},{Pieter G.},{Henry},{Alaina},{Skelton},{Rosalind E.},{Fumagalli},{Mattia},{Momcheva},{Ivelina G.},{Brammer},{Gabriel B.},{Labb{\'e}},{Ivo},{Nelson},{Erica J.},\&{Rigby},{Jane R.}
}]{Whitaker14}
{Whitaker K.~E.}, {et al.} 2014, \apj, 795, 104

\bibitem[{
{Wilkinson}{ et al.}(2017)
{Wilkinson},{Aaron},{Almaini},{Omar},{Chen},{Chian-Chou},{Smail},{Ian},{Arumugam},{Vinodiran},{Blain},{Andrew},{Chapin},{Edward L.},{Chapman},{Scott C.},{Conselice},{Christopher J.},{Cowley},{William I.},{Dunlop},{James S.},{Farrah},{Duncan},{Geach},{James},{Hartley},{William G.},{Ivison},{Rob J.},{Maltby},{David T.},{Micha{\l}owski},{Micha{\l} J.},{Mortlock},{Alice},{Scott},{Douglas},{Simpson},{Chris},{Simpson},{James M.},{van der Werf},{Paul},\&{Wild},{Vivienne}
}]{Wilkinson17}
{Wilkinson A.}, {et al.} 2017, \mnras, 464, 1380

\bibitem[{
{Williams}{ et al.}(2009)
{Williams}, Rik J., {Quadri}, Ryan F., {Franx}, Marijn, {van Dokkum}, Pieter, \&{Labb{\'e}}, Ivo
}]{Williams09}
{Williams R.~J.}, {et al.} 2009, \apj, 691, 1879

\bibitem[{
{Willmer}{ et al.}(2006)
{Willmer}, C.~N.~A., {Faber}, S.~M., {Koo}, D.~C., {Weiner}, B.~J., {Newman}, J.~A., {Coil}, A.~L., {Connolly}, A.~J., {Conroy}, C., {Cooper}, M.~C., {Davis}, M., {Finkbeiner}, D.~P., {Gerke}, B.~F., {Guhathakurta}, P., {Harker}, J., {Kaiser}, N., {Kassin}, S., {Konidaris}, N.~P., {Lin}, L., {Luppino}, G., {Madgwick}, D.~S., {Noeske}, K.~G., {Phillips}, A.~C., \&{Yan}, R.
}]{Willmer06}
{Willmer C.~N.~A.}, {et al.} 2006, \apj, 647, 853

\bibitem[{
{Wyder}{ et al.}(2007)
{Wyder}, Ted K., {Martin}, D. Christopher, {Schiminovich}, David, {Seibert}, Mark, {Budav{\'a}ri}, Tam{\'a}s, {Treyer}, Marie A., {Barlow}, Tom A., {Forster}, Karl, {Friedman}, Peter G., {Morrissey}, Patrick, {Neff}, Susan G., {Small}, Todd, {Bianchi}, Luciana, {Donas}, Jos{\'e}, {Heckman}, Timothy M., {Lee}, Young-Wook, {Madore}, Barry F., {Milliard}, Bruno, {Rich}, R. Michael, {Szalay}, Alex S., {Welsh}, Barry Y., \&{Yi}, Sukyoung K.
}]{Wyder07}
{Wyder T.~K.}, {et al.} 2007, \apjs, 173, 293

\bibitem[{
{Yuan}{ et al.}(2010)
{Yuan}, T. -T., {Kewley}, L.~J., \&{Sanders}, D.~B.
}]{Yuan10}
{Yuan T.-T.}, {Kewley L.~J.}, \& {Sanders D.~B.} 2010, \apj, 709, 884

\bibitem[{
{Zabludoff} \& {Mulchaey}(1998)
{Zabludoff}, Ann I., \&{Mulchaey}, John S.
}]{Zabludoff98}
{Zabludoff A.~I.}, \& {Mulchaey J.~S.} 1998, \apj, 496, 39

\bibitem[{
{Zavala}{ et al.}(2019)
{Zavala}, J.~A.,{Casey}, C.~M.,{Scoville}, N.,{Champagne}, J.~B.,{Chiang}, Y.,{Dannerbauer}, H.,{Drew}, P.,{Fu}, H.,{Spilker}, J.,{Spitler}, L.,{Tran}, K.~V.,{Treister}, E.,\&{Toft}, S.
}]{Zavala19}
{Zavala J.~A}, {et al.} 2019, \apj, 887, 183

\bibitem[{
{Zeballos}{ et al.}(2018)
{Zeballos}, M.,{Aretxaga}, I.,{Hughes}, D.~H.,{Humphrey}, A.,{Wilson}, G.~W.,{Austermann}, J.,{Dunlop}, J.~S.,{Ezawa}, H.,{Ferrusca}, D.,{Hatsukade}, B.,{Ivison}, R.~J.,{Kawabe}, R.,{Kim}, S.,{Kodama}, T.,{Kohno}, K.,{Monta{\~n}a}, A.,{Nakanishi}, K.,{Plionis}, M.,{S{\'a}nchez-Arg{\"u}elles}, D.,{Stevens}, J.~A.,{Tamura}, Y.,{Velazquez}, M.,\&{Yun}, M.~S.
}]{Zeballos18}
{Zeballos M.}, {et al.} 2018, \mnras, 479, 4577

\bibitem[{
{Zhang}{ et al.}(2022)
{Zhang},{Benjamin},{Lee},{Khee-Gan},{Krolewski},{Alex},{Shi},{Jingjing},{Horowitz},{Benjamin},\&{Kooistra},{Robin}
}]{Zhang22}
{Zhang B.}, {et al.} 2022, arXiv:2211.09331





\end{thebibliography}

\end{document}